\documentclass[12pt]{elsarticle}

\usepackage{amsmath,amssymb,amsfonts}
\usepackage{amsthm}
\usepackage[dvipsnames]{xcolor}
\usepackage{dsfont}

\usepackage{caption}
\usepackage{placeins}
\usepackage{rotating}
\usepackage{lscape}
\usepackage{multirow}

\usepackage{helvet} % For sans-serif font

\usepackage{geometry}
\usepackage{tikz}
\usetikzlibrary{%
arrows,%
calc,%
fit,%
patterns,%
plotmarks,%
shapes.geometric,%
shapes.misc,%
shapes.symbols,%
shapes.arrows,%
shapes.callouts,%
shapes.multipart,%
shapes.gates.logic.US,%
shapes.gates.logic.IEC,%
automata,%
er,%
backgrounds,%
chains,%
topaths,%
trees,%
petri,%
mindmap,%
matrix,%
calendar,%
folding,%
fadings,%
through,%
positioning,%
scopes,%
decorations.fractals,%
decorations.shapes,%
decorations.text,%
decorations.pathmorphing,%
decorations.pathreplacing,%
decorations.footprints,%
decorations.markings,%
shadows}
\definecolor{orange}{HTML}{d9944f}
\usetikzlibrary{patterns,arrows,positioning,fit}
\usetikzlibrary{decorations.pathreplacing}
\usetikzlibrary{decorations.pathmorphing}
\usepackage[linesnumbered, ruled, vlined]{algorithm2e}
\usepackage{graphicx,subcaption,lipsum}

\usepackage[hyphens]{url}
\usepackage{doi}
\usepackage{hyperref}
\usepackage[capitalize]{cleveref}
\usepackage{subcaption}
\usepackage{svg}
\usepackage{adjustbox}
\usepackage[linesnumbered,ruled,vlined]{algorithm2e}
\usepackage{hyperref}
\usepackage{microtype} %way better  spacing (especially word breakps)
\RequirePackage[l2tabu, orthodox]{nag}
\usepackage{booktabs} % nicer looking tables

\usepackage{todonotes}
\newboolean{SHOWCOMMENTS}
\setboolean{SHOWCOMMENTS}{true}
\ifthenelse{\boolean{SHOWCOMMENTS}}{
\newcommand{\tdMK}[1]
{\todo[color=blue!25,inline]{\footnotesize{\textbf{ Martin:}} #1}}
}{
\newcommand{\tdMK}[1]{}
}
\newcommand{\MK}[1]{{\color{black}#1}}
\newcommand{\MKSR}[1]{{\color{black}#1}}

\ifthenelse{\boolean{SHOWCOMMENTS}}{
\newcommand{\tdMM}[1]
{\todo[color=cyan!25,inline]{\footnotesize{\textbf{Michael:}} #1}}
}{
\newcommand{\tdMM}[1]{}
}

\ifthenelse{\boolean{SHOWCOMMENTS}}{
\newcommand{\tdDK}[1]
{\todo[color=green!80!gray!25,inline]{\footnotesize{\textbf{David:}} #1}}
}{
\newcommand{\tdDK}[1]{}
}
\newcommand{\DK}[1]{{\color{black}#1}}
\newcommand{\DKSR}[1]{{\color{black}#1}}

\ifthenelse{\boolean{SHOWCOMMENTS}}{
\newcommand{\tdKN}[1]
{\todo[color=yellow!25,inline]{\footnotesize{\textbf{Khoa:}} #1}}
}{
\newcommand{\tdKN}[1]{}
}

\ifthenelse{\boolean{SHOWCOMMENTS}}{
\newcommand{\tdSK}[1]
{\todo[color=teal!20,inline]{\footnotesize{\textbf{Sascha:}} #1}}
}{
\newcommand{\tdSK}[1]{}
}
\newcommand{\SK}[1]{{\color{black}#1}}
\newcommand{\SKSR}[1]{{\color{black}#1}}

\ifthenelse{\boolean{SHOWCOMMENTS}}{
\newcommand{\tdDA}[1]
{\todo[color=orange!20,inline]{\footnotesize{\textbf{Daniel:}} #1}}
}{
\newcommand{\tdDA}[1]{}
}

\let\olditemize\itemize
\renewcommand\itemize{\olditemize\addtolength{\itemsep}{-0.25cm}}%

\journal{Elsevier}%Computers in Biology and Medicine
\begin{document}

\begin{frontmatter}



\title{Agent-based modeling for realistic reproduction of human mobility and contact behavior to evaluate test and isolation strategies 
in epidemic infectious disease spread}

\author[aff1]{David Kerkmann\fnref{fa}}
\ead{david.kerkmann@theoretical-biology.de}

\author[aff2]{Sascha Korf\fnref{fa}}
\ead{sascha.korf@dlr.de}

\author[aff3]{Khoa Nguyen}

\author[aff2]{Daniel Abele}

\author[aff4]{Alain Schengen}

\author[aff2]{Carlotta Gerstein }

\author[aff5]{Jens Henrik G\"obbert}

\author[aff2]{Achim Basermann}

\author[aff2,aff6]{Martin J. Kühn\fnref{ca}}
\ead{martin.kuehn@dlr.de}

\author[aff1]{Michael Meyer-Hermann\fnref{ca}}
\ead{mmh@theoretical-biology.de}

\fntext[fa]{Shared first author in alphabetic order}
\fntext[ca]{Shared corresponding author in alphabetic order}

\affiliation[aff1]{organization={Helmholtz Centre for Infection Research},
            city={Brunswick},
            country={Germany}}
\affiliation[aff2]{organization={Institute of Software Technology, Department of High-Performance Computing, German Aerospace Center},%Department and Organization, 
            city={Cologne},
            country={Germany}}
\affiliation[aff3]{organization={Department of Vulnerabilities and Social Medicine, Unisanté (Center For Primary Care and Public Health)}, city={Lausanne}, country={Switzerland}}
\affiliation[aff4]{organization={Institute for Transport Research, German Aerospace Center},%Department and Organization, 
            city={Berlin},
            country={Germany}}
\affiliation[aff5]{organization={Jülich Supercomputing Centre, Forschungszentrum Jülich},%Department and Organization, 
            city={Jülich},
            country={Germany}}
\affiliation[aff6]{organization={Life and Medical Sciences Institute, University of Bonn},%Department and Organization, 
            city={Bonn},
            country={Germany}}

%% Abstract
\begin{abstract}
Agent-based models have proven to be useful tools in supporting decision-making processes in different application domains. 
The advent of modern computers and supercomputers has enabled these bottom-up approaches to realistically model human mobility and contact behavior.

The COVID-19 pandemic showcased the urgent need for detailed and informative models that can answer research questions on transmission dynamics.
We present a sophisticated agent-based model to simulate the spread of respiratory diseases. 
The model is highly modularized and can be used on various scales, from a small collection of buildings up to cities or countries. 
Although not being the focus of this paper, the model has undergone performance engineering on a single core and provides an efficient intra- and inter-simulation parallelization for time-critical decision-making processes.

In order to allow answering research questions on individual level resolution, nonpharmaceutical intervention strategies such as face masks or venue closures can be implemented for particular locations or agents. In particular, we allow for sophisticated testing and isolation strategies to study the effects of minimal-invasive infectious disease mitigation.

With realistic human mobility patterns for the region of Brunswick, Germany, we study the effects of different interventions between March 1st and May 30, 2021 in the SARS-CoV-2 pandemic. Our analyses suggest that symptom-independent testing has limited impact on the mitigation of disease dynamics if the dark figure in symptomatic cases is high. Furthermore, we found that quarantine length is more important than quarantine efficiency but that, with sufficient symptomatic control, also short quarantines can have a substantial effect.
\end{abstract}


%\item An agent-based model including realistic %individual mobility and transmission of infections
%\item Efficient and scalable implementation for %simulations of several million agents
%\item Our results suggest that symptom-independent 
%testing strategies have limited effect if %symptomatic control is low
%\item Our results suggest that with sufficient
%symptomatic control, even short quarantines  
%have a substantial effect on the mitigation of %disease dynamics

%% Keywords
\begin{keyword}
Agent-based modeling \sep mathematical models \sep testing strategies \sep human mobility \sep infectious diseases



\end{keyword}

\end{frontmatter}



\section{Introduction}\label{sect:introduction}

The recent SARS-CoV-2 pandemic has demonstrated how human societies can be impacted by infectious diseases. While different vaccination strategies have prevented many severe disease outcomes through long-term protection~\cite{BOBROVITZ2023,Stein2023}, nonpharmaceutical interventions (NPIs) like test, trace and isolation (TTI) strategies have been important counteractions against SARS-CoV-2 spread for a long time, but also stayed important when vaccination finally became available. With novel pathogens not or only partially covered by available vaccines, NPIs and TTI strategies need to be studied for Pandemic Preparedness.

Mathematical models are invaluable assets in many application domains, as they allow to proactively study alternative scenarios of future developments. 
For infectious disease dynamics, models based on simple systems of ordinary differential equations (ODEs) are popular tools that allow a quick and easy assessment of situations at hand; see, e.g., several \SKSR{sections} in~\cite{brauer_mathematical_2019}. 
However, these models are often overly simplified and use several unrealistic assumptions. While spatial or demographic stratification can already be resolved within larger ODE systems~\cite{liu_modelling_2022} or recent semi-discrete approaches~\cite{kuhn_assessment_2021,zunker_novel_2024}, unrealistic distribution times~\cite{wearing_appropriate_2005,donofrio_mixed_2004} can be overcome by using a linear chain trick~\cite{Getz2017} or its generalization~\cite{Hurtado2019Generalizations} or even Integro-differential equation-based systems~\cite{Wendler2024IDE}. 
Nevertheless, all these models remain aggregated models and only allow limited heterogeneity in individual reaction to transmission or infection. 
On the other hand, contact-network, agent- (ABMs) or individual-based models allow using arbitrary levels of heterogeneity in individual human behavior and reaction to infection. 

\SK{Very early works laying the foundations to modern agent-based modeling can be traced back to the 1940s to 1950s with John von Neumann's work on self-reproducing automata~\cite{von1966theory}. As computational power increased during the late 20th century, ABMs gained more prominence in various fields, such as in social sciences with Thomas Schelling's segregation model~\cite{schelling1971dynamic} showing the potential of ABMs in the 1970s. As computational capabilities allowed more sophisticated models later on, various ABMs were developed in other domains allowing researchers to explore complex market dynamics, population dynamics, or social phenomena; see~\cite{038f59c7-dd05-312a-82e2-1c92036535ae,doi:10.1073/pnas.082080899,gilbert2000build}. Over the past few decades, increasingly detailed data on individual behaviors and heterogeneous contact patterns have become available. This advancement has enabled the construction of synthetic populations and contact networks, which serve as the foundation for not only our agent-based models but also for most agent-based approaches in infectious disease analysis \cite{RevModPhys.87.925, WANG20161, JUSUP20221, 10.1093/comnet/cnaa041, fan2022epidemics, Eubank2004}.}
\MK{Finally, with the advent of modern computers and supercomputers, highly detailed and parametrized ABMs became feasible to be applied, in particular, to infectious disease spread of different pathogens; see, e.g., }~\cite{collier_parallel_2013,willem_optimizing_2015,venkatramanan_using_2018,bershteyn_implementation_2018,djurdjevac_conrad_human_2018,ferguson2020report,adamik_mitigation_2020,bicher_evaluation_2021,kerr_covasim_2021,muller_predicting_2021,Franek23,Ponge23,ozik2021population,macal2018chisim,gaudou2020comokit,taillandier2019building} \MK{for different developments of the last few decades and recent novel research emerging from the SARS-CoV-2 pandemic.} Note that~\cite{colizza_flu_2010} makes a distinction between \textit{agent-based models} (with movements and interactions of individuals) and \textit{contact-network models} (with a detailed network of social interactions between individuals) which is not made here.

The authors of~\cite{collier_parallel_2013} presented ground-breaking and promising ABMs on HPC infrastructure applied to, e.g., rumor spreading. 
The authors of~\cite{willem_optimizing_2015} presented another very interesting multicore ABM and shed light on many different and costly aspects in agent-based modeling. 
From the geolocations in a 24h interval, the authors of~\cite{venkatramanan_using_2018} derived a social contact network with edge weights given by the duration of contacts. In 2018, the powerful platform EMOD, allowing for environmental, vector, airborne or sexual transmittable diseases, was presented by~\cite{bershteyn_implementation_2018}. In~\cite{djurdjevac_conrad_human_2018}, innovation spreading was modeled through an ABM. 
In~\cite{ferguson2020report}, the authors modified a previously developed simulation tool, originally designed for Influenza studies, to analyze COVID-19 in the UK and the US. Their model includes detailed interactions within households, workplaces, schools, and the wider community, forming a structured contact network that reflects the primary contexts for disease transmission. The level of adherence to interventions, such as case isolation and quarantine, was incorporated into the model with specific compliance rates, influencing the simulation outcomes and reflecting realistic variations in public response. 
The model of~\cite{ozik2021population} was developed using ChiSIM~\cite{macal2018chisim}, a versatile community model initially designed for exploring a range of social dynamics in Chicago, including the transmission of infectious diseases. This model utilizes the Repast ABM framework~\cite{collier_parallel_2013} to facilitate a distributed memory simulation that operates on high-performance computing (HPC) systems. In each simulated hour, agents choose a single activity from their predefined activity schedules to be performed at a specific location. Additionally, agents have the option to alter their behavior and wear masks, practice social distancing, or opting to stay home, which serves as a surrogate for quarantine measures. The calibration of the model, particularly the agents' stay-at-home behavior and the disease model parameters, was aligned with data on hospital bed usage and deaths attributed to COVID-19. The COVID-19 Modeling Kit (COMOKIT)~\cite{gaudou2020comokit} is an agent-based modeling software on the GAMA platform \cite{taillandier2019building}, designed to analyze COVID-19 response strategies in various contexts. COMOKIT combines models of person-to-person and environmental transmission, a model of individual epidemiological status evolution, an agenda-based one\SK{-}hour time step model of human mobility, and an intervention model. It integrates models for transmission, policy, and individual behavior, and is adaptable for different scenarios using minimal initial data. Despite its effectiveness, the current version faces challenges in scaling up for larger populations or fully representing social dynamics in outdoor group activities.
While this list is nonexhaustive, many other authors used ABMs to simulate COVID-19, see, e.g.,~\cite{adamik_mitigation_2020,bicher_evaluation_2021,kerr_covasim_2021,muller_predicting_2021,Franek23,Ponge23,bicker_hybrid_2024}.

The model presented in this study is a trip-based ABM, which means that it explicitly realizes mobility and allows microscopic contact patterns inside locations. Through this approach, NPIs can be realized in a fully natural way by enforcing venues to be closed, capacity-restricted, or by requiring vaccinations or negative tests of any entering agents. In contrast to equation- or mean-field models, it also allows for the study of heterogeneous individual immunity and transmission dynamics and enables settings like face masks, testing, isolation, 
and compliance.
On one hand, our ABM allows for efficient computations of city-scale infectious disease dynamics on a single laptop. On the other hand, our model has been parallelized, both,
on an intra- and inter-simulation level such that large-scale simulations can be efficiently conducted on a\SK{n} HPC infrastructure.

\section{Material and methods}\label{sec:MethodsAndModel}

In this section, we will provide detailed insights in our ABM developed within the MEmilio framework~\cite{memilio121} for high-performance and innovative state-of-the-art infectious disease models. 
We will start from our theoretical understanding of ABMs and a high-level view of the agent-based simulation. Then, major model components will be introduced in separate sections, and finally, we will share our parallelization strategies to enable the execution of large-scale simulation.

\begin{figure}[h!] 
	\centering \includegraphics[width=0.9\linewidth]{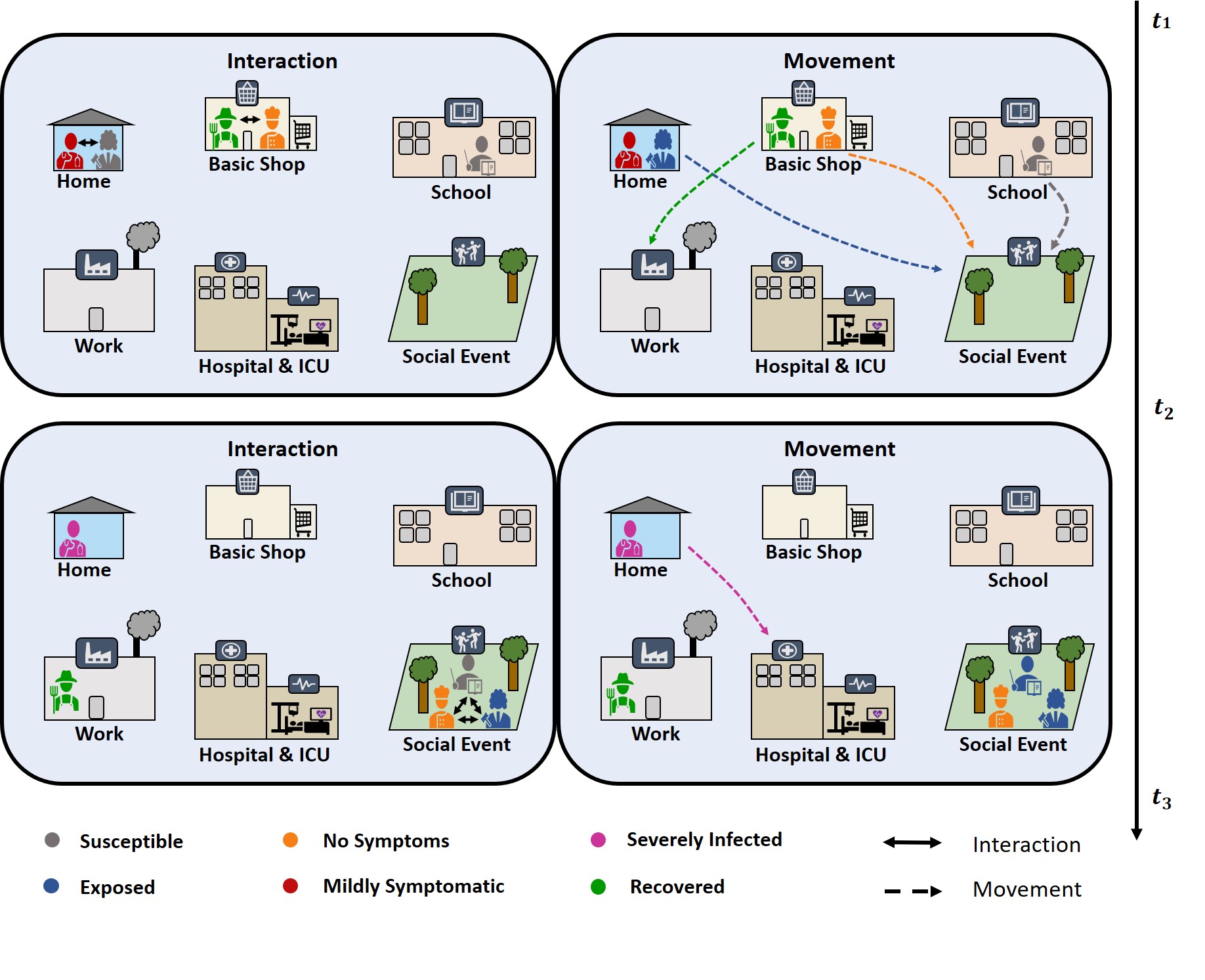} 
	\caption{\textbf{Two exemplary time steps of the ABM simulation with five agents.} From top left to bottom right, persons interact and potentially infect each other at a location. Then, agents potentially move to other location types, e.g., from a supermarket to work, or \SK{to the hospital} if they have severe symptoms. After a  time step, the infection state can progress\SK{, e.g., from mildly infected to severely infected.}}
	\label{fig:Modelflow}
\end{figure}

\DK{Furthermore, many technical details and formulas are presented in~\cref{sec:transmission}, \cref{sec:testing}, \cref{sec:rng} and \cref{sec:parallel}. Although these details are important to fully understand the entire model and its capabilities, a general understanding can be obtained without the technical details, in particular through the support of \cref{code:simulation_1} to \cref{code:simulation_interaction} and \cref{fig:Modelflow} to \cref{fig:test_strategies}.}

In our understanding, an \emph{agent-based model (ABM)} consists of a finite number of \emph{agents} and an \emph{environment} that hosts agents in which agents act and react to other agents.
Furthermore,
\begin{itemize}
    \item an agent is characterized by a finite number of \emph{features} that determine its state,
    \item an agent interacts with other agents and their joint environment according to \emph{interaction rules},
    \item the state of an agent or the environment changes through interactions or with time.
\end{itemize}
Several items, such as the environment or the interaction rules, which depend on the respective realization, will be introduced in the following subsections.

In~\cref{code:simulation_1}, we provide the highest level view on the agent-based simulation process on $[t_0,t_{\max}]$ with step width $\Delta t$. 

\begin{algorithm}
\caption{\textbf{Trip-based agent-based simulation}}
\label{code:simulation_1}
$t \leftarrow t_0\in\mathbb{R}$ \\
\While{$t  \leq t_{\max}$ } {
    \For{each location}{
        \textbf{Execute agents' interactions}
    }
    \For{each agent}{
        \textbf{Perform individual movement}
        }
    $t \leftarrow t+\Delta t $  
}
\end{algorithm}

There are usually two different conceptualizations for what is often denoted an \textit{ABM}.
On the one hand, a graph is realized with the agents as nodes, edges as particular contacts and edge weights as the number, duration, or probability of transmission, as, e.g., in~\cite{willem_optimizing_2015,kerr_covasim_2021}. On the other hand, trip-based ABMs explicitly move agents to locations (which could be considered as the nodes of a graph). Note that~\cite{colizza_flu_2010} refer only to the latter as \textit{agent-based models} while the prior are denoted \textit{contact-network models}.
In both cases, an agent can represent a single individual but in simplifications also a group of people in the real world; cf.~\cite{kerr_covasim_2021}. 

Our implementation of the trip-based ABM has the advantage that the amount of agent-to-agent interactions gets defined by the activities in the locations where the trips go to, e.g., school or work. Agents can only interact with other agents at the same location during a time step. This means that the potential quadratic complexity $\mathcal{O}(n_{\textrm a}^2)$ in the number of agents $n_{\textrm a}$ roughly reduces to $\mathcal{O}(n_{\textrm a})$, or explicitly to $\mathcal{O}(n_{\textrm l}*m^2)$, where $n_{\textrm l}$ is the number of locations and $m$ the maximum number of agents at any location. In~\cref{fig:linear_time_mem_scaling}, we see that computation time and memory usage scale linearly with the number of agents.

Two exemplary time steps of the simulation are depicted in~\cref{fig:Modelflow}. Agents are defined in the following~\cref{sec:agents}, then locations in~\cref{sec:locs}, and finally mobility in~\cref{sec:mobility}.

\subsection{Agents}\label{sec:agents}

As age has early been detected as key feature to determine disease severity~\cite{verity_estimates_2020}, the model is defined for a preselected set of age groups. All agents are assigned a particular age group and many parameters have to be specified for all age groups. Furthermore, immunity and immunity waning are  
important individual properties~\cite{nordstrom_risk_2022}
so that immune histories of infections and vaccinations have to be stored for all agents.
To determine ongoing disease dynamics in the simulation, the agents' locations, potential infection and quarantining, positive and negative tests as well as mask wearing and compliance are of high importance.

In our model, an agent represents an individual person that possesses the following features:
\begin{itemize}
\item age group,
\item current location and the %mmh spent time 
time spent %mmh
at this location,
\item a set of assigned locations, such that the 
    person always goes to the same workplace or school and returns to the same home,
\item %mmh all past 
history of %mmh
vaccinations (with potentially different vaccines),
\item %mmh all past and 
history of and %mmh
current infections (with potentially different virus variants),
\item the point in time quarantine started if the person is quarantined,
\item the point in time of the last negative test,
\item the type of mask the person usually uses and whether the mask is currently worn,
\item compliance to mask mandates (from voluntary to full refusal),
\end{itemize}
and a counter for the person-specific random number generator that triggers stochastic events. 

\subsection{Locations}\label{sec:locs}
Locations are buildings, spaces or artificial places for the corresponding activities. All locations are informed of the currently present agents to manage interventions such as capacity restrictions and to correctly perform interactions. Locations furthermore have the following features:
\begin{itemize}
\item a type of location (i.e. \textit{Home}, \textit{School}, \textit{Work}, \textit{SocialEvent}, \textit{Shopping}, \textit{Hospital}, \textit{ICU}, and \textit{Cemetery}),
\item a maximum number of contacts and contact rates,
\item a reduction factor for the likelihood of entering the location (when restrictions are in place) and information on location-specific restrictions such as the maximum number of persons allowed to enter the location,
\item and  
whether it is required to wear a mask and which type of mask is mandatory, e.g., FFP2 versus surgical.
\end{itemize}
Users can add more location types by simply extending the list of locations in the code base.

\subsection{Mobility}\label{sec:mobility}
As mentioned in~\cref{code:simulation_1}, line 5, agents perform individual trips to regularly change their location, 
i.e., to go from home to work, to go shopping after work or to go to the hospital if an infection turns out to be severe. The last example is part of our core mobility rule set. This set describes movement depending on the agents' infection state, takes precedence over all other mobility actions and consists of rules to go to and return from medical care facilities. Besides this core mobility, we provide two different realizations of natural mobility. In a basic implementation, we provide \textit{extended mobility rules} for individual agents, according to, e.g., their age 
(see~\cref{sec:mobility_rules}). If individual mobility or trip data is available, we provide an interface to initialize all individuals with individual trip chains that are executed throughout the simulation (see~\cref{sec:trips}).
For settings where individual trip or mobility data is not available, the extended mobility rules always serve as a fallback implementation.

\subsubsection{Mobility rules}\label{sec:mobility_rules}

Mobility rules provide a mechanistic mobility model that realizes mobility depending on the age of the agent, the day of the week and the time of the day.

On the one hand, we have our core set of infection state dependent mobility rules which force agents to quarantine if tested positive, to go to a hospital if severely infected, to move to ICU if critically infected and to go home if recovered. Dead agents are removed from the simulation and stored in a cemetery object for post-simulation analysis. 

On the other hand, we have the extended set adding natural mobility rules which force agents of particular age to go to school between Monday and Friday while agents of other age groups go to work on these same days. The particular age (groups) and times to go to or come from these locations can be set by the user. Similar rules are available for shopping or social events.

Our current implementation o\SK{f} mobility rules is rather basic. Depending on the application, customs and cultural or regional practices like Easter holidays, Christmas or particular New Year's events, more advanced schemes have to be provided by the user.

\subsubsection{Trip data set-based mobility}\label{sec:trips}

To allow for a more realistic and data-driven mobility approach, we allow for the initialization of agents with particular trip chains that are then conducted repeatedly throughout the simulation. To the trips, we add all infection state dependent mobility rules.

The generation of our trip data set is based on a two-stage process. We started from the macroscopic transport demand model DEMO~\cite{WINKLER2020102476} and a population upscaled from the MiD 2017 ("Mobilität in Deutschland") survey for Germany~\cite{nobis2018mobilitat}, which was spatially distributed according to the BKG household dataset (households, inhabitants, federal government) of reference year 2020~\cite{bkg_haushalte_nodate}. In the first step of the process, the trip chains between the DEMO traffic cells were generated based on the daily schedules of the MiD population~\cite{dlr188443}. In the second step of the process, corresponding locations were assigned within the target traffic cells given by the DEMO model. 
The locations were previously extracted from OpenStreetMap and attributed with activities according to their attributes/metadata (key/value pairs)~\cite{MALKUS2024420}. We provide
the data set at~\cite{schengen_2024_13318436}. Note that we currently do not simulate transmission in different transport modes.

With the provided description, we can detail line 6 of~\cref{code:simulation_1} in~\cref{code:simulation_movements}. Details on testing, masks, restrictions, and compliance as of line 6 of~\cref{code:simulation_movements} will be given in~\cref{sec:interventions}.

\begin{algorithm}
\caption{\textbf{Perform individual movement}\\
(Detailed algorithm for line 6 of \cref{code:simulation_1})}
\label{code:simulation_movements}
        \eIf{agent needs to move due to infection severity or test result}{
            Move agent to hospital, ICU, cemetery, or quarantine\;
        }{
            Check individual trips\;
            \If{trip is scheduled in $[t, t + \Delta t]$}{
                \If{agent complies with testing, mask and capacity restrictions}{
                    Move agent to new location\;
                }
            }
        }
\end{algorithm}

\subsection{Contacts}\label{sec:contacts}

The mobility and presence of agents at locations as described in the previous sections define sets of potential contacts of the agents. One option to \SK{model} close contacts, is the use of human locomotion models inside locations, see, e.g.,~\cite{rahn_modelling_2022,rahn_toward_2024}. In general, the average number of daily contacts individuals have inside an 
ABM 
should be in line with aggregated numbers as obtained by~\cite{mossong_polymod,fumanelli_inferring_2012,prem_projecting_2017} and used in ODE-based models. However, as these numbers are reported on an aggregated level, it is impossible to obtain a unique disaggregated, individual number of contacts per location. Furthermore, a trip-based ABM will often yield particular situations in which the average number of contacts cannot be reached by individual agents. On the other hand, individual-based data sets cannot be obtained due to data protection regulations.

An approach to the described problem is to formulate a global optimization problem (with constraints). For small problems with a handful of locations, optimal solutions could be found. For larger problems, additional complexities and high run-times came into play. In~\cref{sec:transmission}, we present how we model the transmission process without using a close contact implementation.

% \subsection{Contact and Movement Model}\label{sec:mobility} % Khoa 

% The model computes the first 3 layers (trip generation, trip distribution and modal split) of the 4-step transport model. 
% The input for such a model consists of a synthetic population grouped into households, capacity constrained locations for activities, daily activities routines and choice models for location and mode choice. The population is generated by an in-house application called SYNTHESIZER \cite{SIMUL}. 
% Inside TAPAS, the population is represented as agents that are attributed with socio-demographic parameters such as age, gender, labor status (part/full-time working), mobility budgets and options like owner of a monthly public transport ticket that influence the choices of agents \cite{10.1007/s00779-017-1031-3}. 
% As for the Brunswick use case, an extract from a larger region model in Lower Saxony spanning  from Hannover over Hildesheim and Brunswick to Wolfsburg has been used as well as the synthetic population that was assigned in Brunswick. 

\subsection{Transmission and infection model}\label{sec:infection_model}
At the heart of the ABM lies the transmission and infection, i.e., course of the disease, model. 
Taking into account each agent's age and immune status due to previous infections and vaccinations, we model the transmission risk and disease progression individually for each agent. In the following subsections, we first introduce viral load and symptom states before we explain how we realize the transmission process.

\DK{As this section is of high technical detail, we provide a quick summary \SK{beforehand} which enables a quick, shallow understanding of our infection and transmission model: We describe each agent's infectiousness time-dependent, beginning at zero at infection transmission, rising during the infection until the peak level of viral load, and then slowly dropping again to zero until the virus is cleared from the host. During a contact with an infected agent, this time-dependent infectiousness is evaluated, and the chance of infection is derived, taking into account further properties of both the transmitting and receiving agent, such as the use of masks and an individual protection level from prior infections or vaccinations. In addition, each agent undergoes a series of symptomatic states, beginning at the time of positive virus transmission, and ending in a recovered or dead state. This state directly influences whether and when the agent visits the hospital or is tested.}
\SKSR{Building on this quick overview, we now explain the realization of the infection and transmission model in detail, starting with the twofold representation of an infection through viral load and infection states in \cref{sec:vlandis} and how agents infect each other in \cref{sec:transmission}.}

\subsubsection{Viral load and infection states}\label{sec:vlandis}
We consider two main representatives of an infection, \textit{viral load} and (symptomatic) \textit{infection states}.
The viral load time course of an infected agent $p$ describes the amount of virus particles carried by the host. It is modeled by a continuous function in time. As observed in~\cite{jones_estimating_2021}, viral load time courses consist of a rapid exponential increase \SK{in} virus particles until a peak level, followed by a shallow exponential decrease until the virus is cleared by the immune system.
Mathematically, we model the time of peak viral load as $t_{\textrm{P,p}} = t_{\textrm{T,p}} + \frac{v_{\textrm{P,p}}}{v_{\textrm{I,p}}}$, where $t_{\textrm{T,p}}$ is the time of transmission and $v_{\textrm{I,p}},v_{\textrm{P,p}}>0$ are the viral load increase rate and peak, respectively.
The logarithmic viral load $v_p$ can be described as
\begin{equation}
    \label{eqn:viral_load}
    v_{\textrm p}(t):=
    \begin{cases}
        (t-t_{\textrm{T,p}}) v_{\textrm{I,p}} & t_{\textrm{T,p}} \leq t \leq t_{\textrm{P,p}} \\
        v_{\textrm{P,p}} + (t-t_{\textrm{P,p}})v_{\textrm{D,p}} & t_{\textrm{P,p}} \leq t \leq t_{\textrm{C,p}}, \\
    \end{cases}
\end{equation}
\begin{figure} 
	\centering 
    \includegraphics[width=0.48\linewidth]{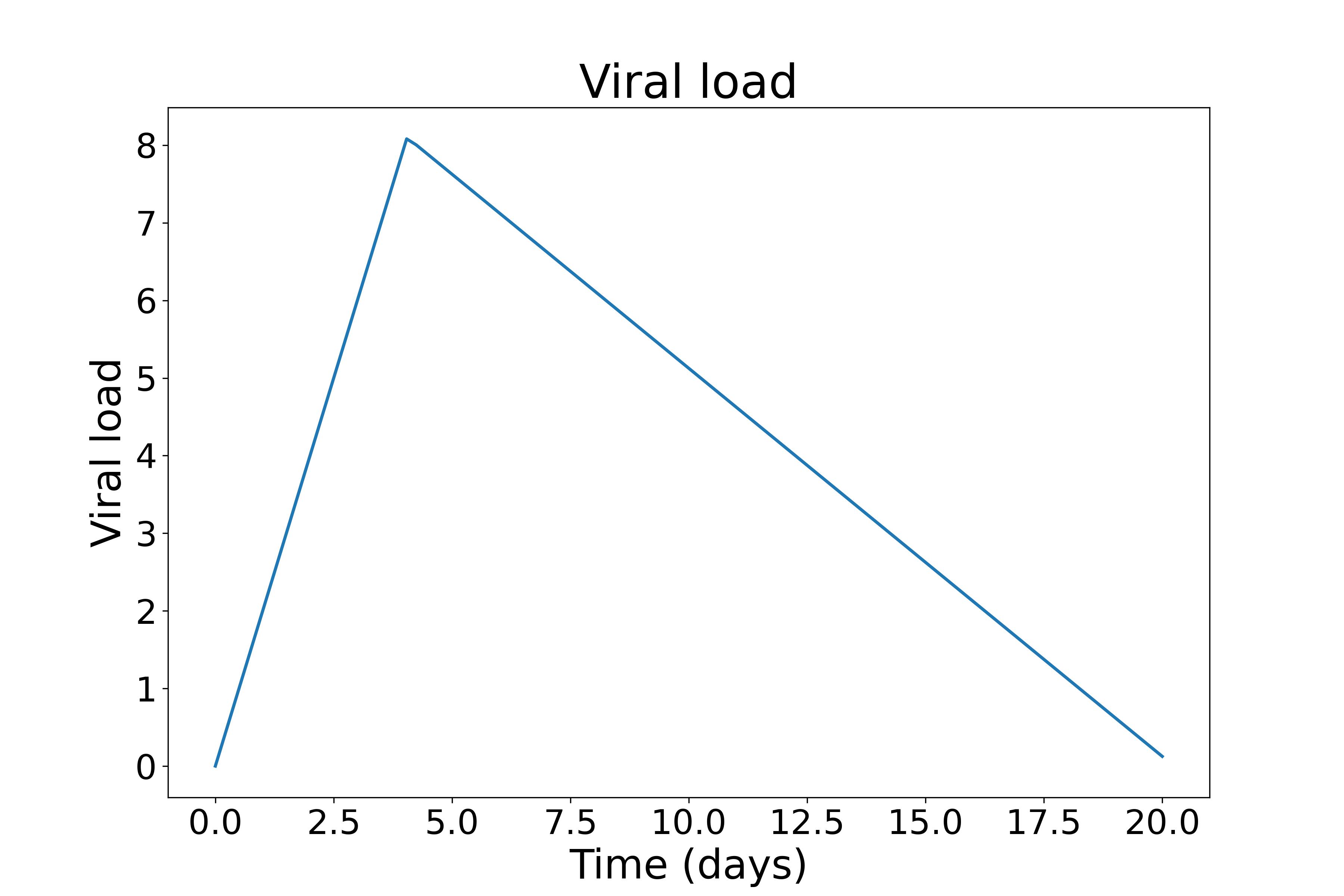} 
\hspace*{0.25cm}
	\includegraphics[width=0.48\linewidth]{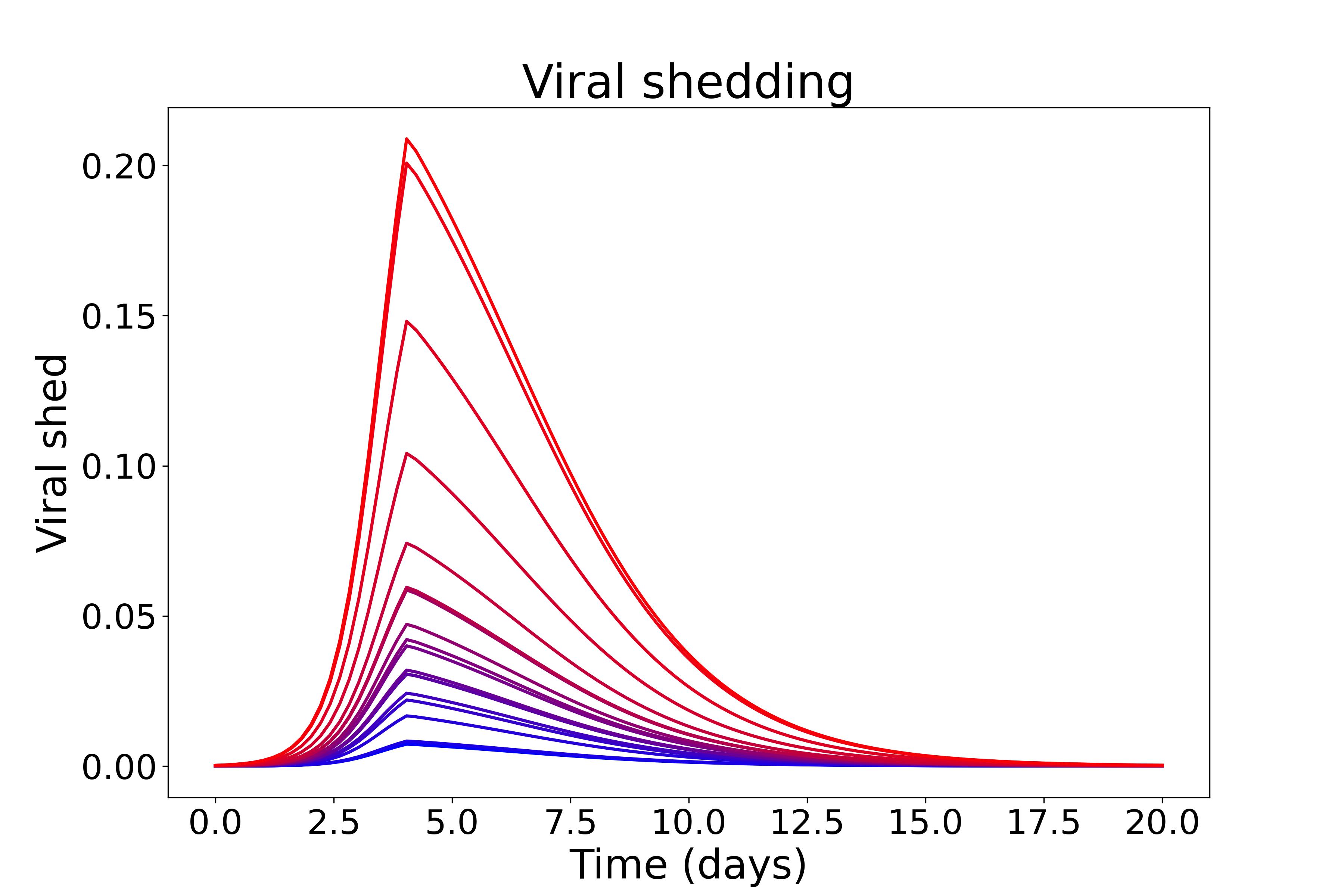} 
	\caption{\textbf{\SK{Course of }viral load (left) and different, corresponding viral sheds (right) \SK{over 20 days}}. The presented viral sheds are given for varying factor $s_{\textrm {f,p}}$ in~\eqref{eqn:viral_shed}, ranging from 0.01 (blue) to 0.28 (red).}
	\label{fig:viral_load}
\end{figure}
with $v_{\textrm {D,p}}<0$ the viral load decrease rate and $t_{\textrm {C,p}}$ as the time the virus is basically cleared by the host's immune system, i.e., $v_{\textrm p}(t_{\textrm {C,p}}) = 0$ and neglect logarithmic viral load $v_{\textrm p} < 0$. For simplicity, from now on, the logarithmic viral load is referred to as viral load, only.

Second, we model infection states as a proxy for the symptoms during an infection. 
Upon transmission, an agent is considered exposed, and will subsequently traverse one path of the possible symptomatic infection courses as shown in~\cref{fig:InfectionStates}. 
The duration in a state as well as which state is taken next is drawn with the help of a random number generator, always taking into account age group- and virus-specific transmission probabilities and duration of stay parameters.

While an early systematic review~\cite{rao_systematic_2020} and, e.g.,~\cite{waller_viral_2023} found higher SARS-CoV-2 viral loads associated with more severe COVID-19, a more recent systematic review~\cite{dadras_relationship_2022} was inconclusive and found contradictory studies on the relationship between viral load and COVID‐19 severity. Accordingly, we model symptoms independently of the viral load.
However, the time of symptom onset was observed to be close to the time of peak viral load~\cite{walsh_duration_2020,puhach_sars-cov-2_2023}. When setting parameters, this is taken into account.

\begin{figure} [h!]
	 \centering
 \begin{tikzpicture}[node distance=4cm, every node/.style={scale=0.9}]
\tikzstyle{block} = [rectangle, minimum width=3cm, minimum height=1cm, text centered, draw=red, line width=0.5mm, fill=gray!30]
\tikzstyle{blockgray} = [rectangle,rounded corners=.35cm, minimum width=3cm, minimum height=1cm, text centered, draw=gray, line width=0.5mm, fill=gray!30]
\tikzstyle{arrow} = [thick,->,>=stealth, line width=0.5mm]
\tikzstyle{dashedblock} = [rectangle, dashed, minimum width=3cm, line width=0.5mm, minimum height=1cm, text centered, draw=red, fill=gray!30]
% Erste Reihe
\node (susceptible) [blockgray, fill={rgb,255:red,176; green,176; blue,176}] {Susceptible}; % Grau
\node (exposed) [blockgray, fill={rgb,255:red,217; green,232; blue,252}, right of=susceptible] {Exposed}; % Hellblau
\node (no_symptoms) [block, fill={rgb,255:red,245; green,134; blue,52}, right of=exposed] {No Symptoms}; % Orange
\node (mild) [dashedblock, fill={rgb,255:red,192; green,0; blue,0}, right of=no_symptoms] {\color{white}\textbf{Mild}}; % Rot

% Spalten
\node (severe) [dashedblock, fill={rgb,255:red,204; green,51; blue,153}, below of=mild, yshift=2cm] {\color{white}\textbf{Severe}}; % Magenta
\node (critical) [dashedblock, fill={rgb,255:red,112; green,68; blue,160}, below of=severe, yshift=2cm] {\color{white}\textbf{Critical}}; % Lila

\node (recovered) [blockgray, fill={rgb,255:red,40; green,167; blue,69}, left of=severe] {\color{white}\textbf{Recovered}}; % Grün
\node (dead) [blockgray, fill={rgb,255:red,77; green,77; blue,77}, left of=critical] {\color{white}\textbf{Dead}}; % Dunkelgrau

% Pfeile
\draw [arrow] (susceptible) -- (exposed);
\draw [arrow] (exposed) -- (no_symptoms);
\draw [arrow] (no_symptoms) -- (mild);
\draw [arrow] (mild) -- (severe);
\draw [arrow] (severe) -- (critical);

% Verbindungen zu den Endzuständen
\draw [arrow] (no_symptoms) -- (recovered);
\draw [arrow] (mild) -- (recovered);
\draw [arrow] (severe) -- (recovered);
\draw [arrow] (critical) -- (recovered);
\draw [arrow] (critical) -- (dead);
\end{tikzpicture}
	\caption{\textbf{Infection state transition model.} 
    \SK{One-way arrows represent the possible transitions from one infection state to another.}
 Rounded gray borders indicate noninfectious states
 while red angular borders indicate symptom states that are used for infectious agents only. Orange to violet colors indicate the symptomatic severity 
 and dashed borders indicate if a person is considered as symptomatic (for testing schemes).}
	\label{fig:InfectionStates}
\end{figure}

\subsubsection{Transmissions}\label{sec:transmission}
% An agent is considered to be infectious, when its viral load is positive and, by default, the infection state is either No Symptoms, Mild, Severe or Critical, compare to \autoref{fig:InfectionStates}. 
% In particular an agent who is exposed or recovered is considered not infectious for other agents.
%mmh In our model, we 
\SKSR{In order to model the transmission of an infection between agents,} we consider the viral shed as the proxy of the infectiousness of an agent. This approach is advantageous because viral shedding directly reflects the amount of virus being released by an individual, which has been shown to have a large impact on transmission potential~\cite{marc_quantifying_2021}. Unlike symptom-based measures, which may lag behind or be absent in asymptomatic cases, viral shedding captures infectiousness even before symptoms appear or in cases where individuals remain asymptomatic throughout the infection~\cite{he_temporal_2020, walsh_duration_2020}. This proxy also allows for a more granular and dynamic assessment of infectiousness over time. As studies~\cite{walsh_duration_2020,puhach_sars-cov-2_2023} have observed that viral load peaks close to symptom onset, symptom states alone would underestimate the transmission risk during the early, highly infectious phase~\cite{jones_estimating_2021, ke_daily_2022}.

In~\cite{ke_daily_2022}, the level of infectious virus shed was related to the viral load. 
Similarly, and according to~\cite{jones_estimating_2021,jones_virologycharitesars-cov-2-vl-paper_2021}, we compute the rate of virus particles shed by an agent through the nonlinear relation
\begin{align}
    \label{eqn:viral_shed}
    % Dimensions: alpha: dimensionless, beta: 1/log viral load, s_i: shed/day
    s_{\textrm p}(t) = \frac{s_{\textrm {f,p}}}{1 + \exp(- (\alpha + \beta v_{\textrm p}(t)))},
\end{align}
where $s_{\textrm {f,p}}$ is the personal viral shed factor for the agent $p$. The viral shed factor is drawn randomly from a probability distribution for each agent and can vary greatly, c.f.~\cite[Fig.~3]{ke_daily_2022}; also see~\cref{fig:viral_load}.
\noindent Similarly to the viral load above, we neglect any viral shed outside of the time interval $[t_{\textrm{T,p}},t_{\textrm {C,p}}]$. Furthermore, we consider only agents which are not Susceptible, Exposed, Recovered and Dead to be infectious. For instance, this means that potential viral loads larger zero are cut on symptomatic recovery. As this leads to a potential cutoff of $s_{\textrm p}(t)$, this needs to be taken into account, when determining $\alpha$ and $\beta$. In our simulations, cutting viral loads for recovered people happens mostly when the viral shed is already low, thus leading to a minimal loss of total infectiousness.

As explained in~\cref{sec:contacts}, we do not model close contacts explicitly but want our results to be in line with the number of contacts as obtained by surveys such as~\cite{mossong_polymod}. As a basis, we take the location-specific and age-stratified contact matrices as described in~\cite{kuhn_assessment_2021}. These contact rates are scaled to the average duration of stay in each type of location, as contacts can only be made when agents are present at the respective location~\cite{destatis_time}. For locations \textit{Social Event} and \textit{Shopping}, we disassemble the \textit{Other} contact matrix, assuming a 50~\% higher contact rate at social events than in shopping situations.

\begin{algorithm}
\caption{\textbf{Execute agents’ interactions}\\
(Detailed algorithm for line 4 of \cref{code:simulation_1})}
\label{code:simulation_interaction}
        Compute exposure rate through aggregated viral shed of all \\ 
        \hskip 2.0em infected persons in the location\;
        \For{each agent in the location}{
            Determine if transmission occurs using exposure rate, \\ 
            \hskip 2.0em contact matrix, and personal protection factors (e.g.,\\
            \hskip 2.0em  vaccinations, masks)\;
            \If{transmission occurs}{
                Draw new course of infection
            }
        }
\end{algorithm}

To address the issue of aggregated information on contact data, we proceed as shown in~\cref{code:simulation_interaction}. First, we iterate over all infected agents only and, for each age group, compute the average viral shed rate per location during a time step. Then, we iterate over all susceptible individuals at the same location and compute an agent-specific infectious viral shed rate that is received. Here, the received shed is obtained from the emitted shed, corrected through the agent-specific protection factors like masks and prior immunity. In detail, we take the steps as described in the following.

First, we compute the average viral shed rate given by age group $i$, obtained from the individually corrected viral shed rates, during (simulation) time $[t, t + \Delta t)$ as
\begin{align}
    \label{eqn:aggregated_reduced_viral_shed}
    \widetilde{s}_i(t) := \frac{\sum\limits_{\widetilde{p}\in\Omega_i\,:\,v_{\widetilde{\textrm p}}(t)>0} (1-q_{\textrm e, \widetilde{\textrm p}}(t))\, (1-m_{\widetilde{\textrm p}}(t))\, s_{\widetilde{\textrm p}}\Big(t+\frac{\Delta t}{2}\Big)}{|\Omega_{i}|},
\end{align}
where $\Omega_{i}$ is the set of all agents of age group $i$ at the considered location 
and time $t$, $q_{\textrm e, \widetilde{\textrm p}}(t)$ realizes a reduction of the viral shed rate through potential quarantine precautions and distancing and $m_{\widetilde{\textrm p}}(t)$ realizes mask wearing of the infected agent $\widetilde{p}$. 
Note that $q_{\textrm e, \widetilde{\textrm p}}$ and $m_{\widetilde{\textrm p}}$ are piecewise constant on the intervals $[t_{\textrm 0}, t_{\textrm 1}), \ldots, [t_{\max-1}, t_{\max})$. For quarantine, we have 
$0<q_{\textrm e, \widetilde{\textrm p}}\leq 1$ and $q_{\textrm e, \widetilde{\textrm p}}=0$ if there is no quarantine as well as $0<m_{\widetilde{\textrm p}}\leq1$ if a mask is worn and $m_{\widetilde{\textrm p}}=0$ otherwise.
As the viral shed rate is modeled continuously, we use a midpoint rule and evaluate $s_{\textrm p}(t+\Delta t/2)$ to have a more accurate average shed rate for the interval $[t, t+\Delta t)$.

Then, we compute the exposure rate ${e}_j(t)$ for each receiving agent $p$ in age group $j$ 
and time step $[t, t+\Delta t)$ via
\begin{align}
    \label{eqn:exposure_rate}
     {e}_{j}(t) = \sum_{i}   \psi(t)\,r(t)\,c(j, i)\widetilde{s}_i\left(t\right).
\end{align}
Here, $c(j, i)$ is the contact matrix entry of age group $j$ to meet age group $i$ as explained above.
The parameter $\psi(t)$ denotes a seasonality parameter that adjusts contact nature changes due to seasonality and weather conditions. $0<r(t)\leq1$ is a parameter for modeling overall contact reduction, equally for all agents, through NPIs and realizes social distancing or, implicitly, the replacing of large social events through smaller social events.

Finally, to turn the received exposure rate into successful transmissions, we model the resulting individual infection rate $\tau_{\textrm p}$ for the agent $p$ as
\begin{align}
    \label{eqn:infection_rate_indiv}
    \tau_{\textrm p}(t) = l\left({e}_{j}(t)\, (1-m_{\textrm p}(t))\right).
\end{align}
Here, $m_{\textrm p}(t)$ is the protection of a particular 
mask worn by the receiver.

We further use the function $l$ as a function mapping the received and protection-corrected exposure rate to the resulting infection rate.
We assume a linear relationship, 
i.e., $l(x)=\lambda x$ with $\lambda>0$ and determine this factor in the fitting process of the simulation.
Then, $\tau_{\textrm p}$ is the infection rate of the agent. Let further $X \sim \exp(\tau_{\textrm p})$ be a random variable and $x$ a realization of $X$. Then a transmission of the virus happens in the time period $[t, t + \Delta t)$ if $x \leq \Delta t$.

As an example, four infected agents (B, C, D, and E) are in a workplace with Agent A, who is susceptible. Each infected agent has the same age but a different viral shedding rate and level of protection from masks, as shown in \cref{tab:ex_individual_infection_rate}. The average viral shed rate at the location is calculated based on the viral shedding rates and the protective measures applied by each infected agent. The average exposure rate for this location is $0.178$.

\begin{table}[ht]
\centering
\caption{\textbf{Example of Factors Affecting Average Viral Shed Rate.}}
\label{tab:ex_individual_infection_rate}
\begin{adjustbox}{width=\textwidth}
\begin{tabular}{@{}lcccc@{}}
\toprule
\textbf{Agent} & \textbf{Viral Shedding Rate ($s_{\widetilde{\textrm p}}(t)$)} & \textbf{Mask Protection ($m_{\widetilde{\textrm p}}(t)$)} & \textbf{Average Viral Shed Rate Contribution} \\ 
\midrule
\textbf{Agent B} & 0.7 & 0.8 & $1 \cdot (1 - 0.8) \cdot 0.7 = 0.14$ \\ 
\textbf{Agent C} & 0.9 & 0.5 & $1 \cdot (1 - 0.5) \cdot 0.9 = 0.45$ \\ 
\textbf{Agent D} & 0.6 & 0.9 & $1 \cdot (1 - 0.9) \cdot 0.6 = 0.06$ \\ 
\textbf{Agent E} & 0.8 & 0.7 & $1 \cdot (1 - 0.7) \cdot 0.8 = 0.24$ \\ 
\midrule
\textbf{Total} & - & - & $0.14 + 0.45 + 0.06 + 0.24 = 0.89$ \\ 
\midrule
\textbf{Average Viral Shed Rate} & & & $\frac{0.89}{5} = 0.178$ \\ 
%mmh \textbf{Average Viral Shed Rate} & \multicolumn{3}{c}{$\frac{0.89}{5} = 0.178$} \\ 
\bottomrule
\end{tabular}
\end{adjustbox}
\end{table}

Agent~A's exposure rate is further adjusted by seasonality and NPIs, resulting in a final exposure rate of $0.128$ for Agent~A. Given that Agent~A is wearing a mask that provides $m_{\textrm p}=85$~\% protection (\(1-m_{\textrm p}(t) = 0.15\)), the individual infection rate for Agent~A is calculated as $0.038$, as detailed in \cref{tab:ex_agent_infection_rate}. This infection rate will be used to determine the likelihood that Agent~A contracts the virus in this specific time step. \SKSR{The implementation of above mentioned NPIs will be explained in detail in the following section.} 

\begin{table}[ht]
\centering
\caption{\textbf{Calculation of Agent A's Infection Rate.}}
\label{tab:ex_agent_infection_rate}
\begin{adjustbox}{width=\textwidth}
\begin{tabular}{@{}lcc@{}}
\toprule
\textbf{Factor}                        & \textbf{Value}                 & \textbf{Calculation (Formula)}  \\ 
\midrule
\textbf{Average Viral Shed Rate}         & 0.178                          & Given \\ 
\textbf{Seasonality Factor ($\psi(t)$)}   & 1.0                            & Given \\ 
\textbf{Contact Reduction ($r(t)$)}    & 0.9                            & Given \\ 
\textbf{Contact Matrix ($c(j,i)$)}     & 0.8                            & Given \\ 
\textbf{Final Exposure Rate ($e_j(t)$)} & 0.128                          & $0.178 \cdot 1.0 \cdot 0.9 \cdot 0.8 = 0.128$ \\ 
\textbf{Mask Protection for Agent A ($m_{\textrm p}(t)$)}   & 0.85 (85~\% protection)         & Given \\ 
\textbf{Linear Infection Coefficient ($\lambda$)} & 2.0                     & Given \\ 
\textbf{Infection Rate for Agent A ($\tau_{\textrm p}(t)$)} & 0.015                   & $2.0 \cdot 0.128 \cdot (1 - 0.85) = 0.038$ \\ 
\bottomrule
\end{tabular}
\end{adjustbox}
\end{table}

\subsection{Nonpharmaceutical interventions}\label{sec:interventions}

To assess retrospective situations and to evaluate potential outcomes of different mitigation strategies, we provide the possibility to implement several NPIs in our ABM. At the center of this study are advanced testing and isolation strategies that are detailed in the next subsection.

\subsubsection{Testing and Isolation}\label{sec:testing} 

\begin{figure} 
	\centering \includegraphics[width=0.75\linewidth]{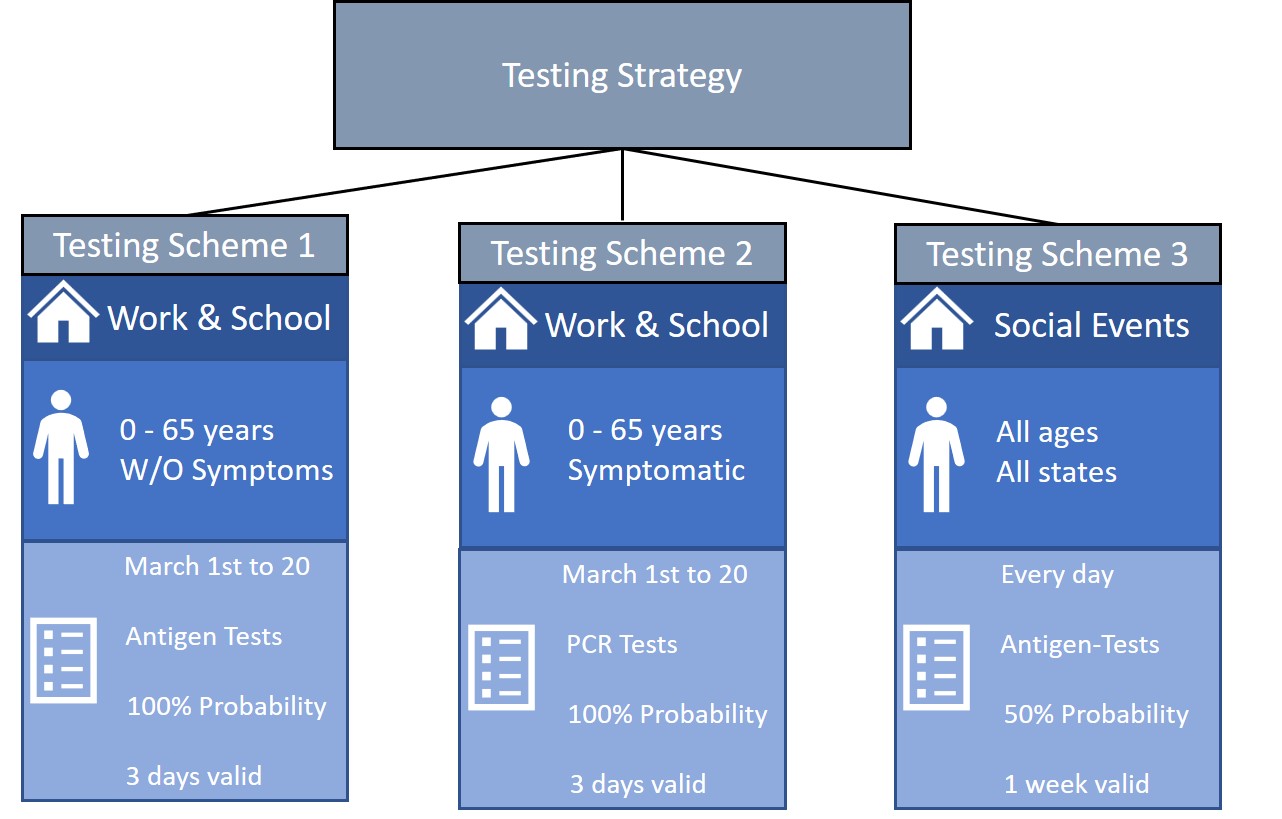} 
	\caption{\textbf{An exemplary Testing Strategy with three different Testing Schemes and different Testing Criteria (middle part, blue).} \SK{We define Testing Schemes for specific locations or, more generally, for different location types (top). Each time an agent enters one of these locations, the Testing Criteria are evaluated (middle). If the criteria are met, the agent undergoes testing with a specified probability and test type. The resulting test remains valid for a predetermined duration. Additionally, the testing strategy can be configured to be active only for a specified time period (bottom).}}
	\label{fig:test_strategies}
\end{figure}

\begin{figure} 
	\centering \includegraphics[width=0.75\linewidth]{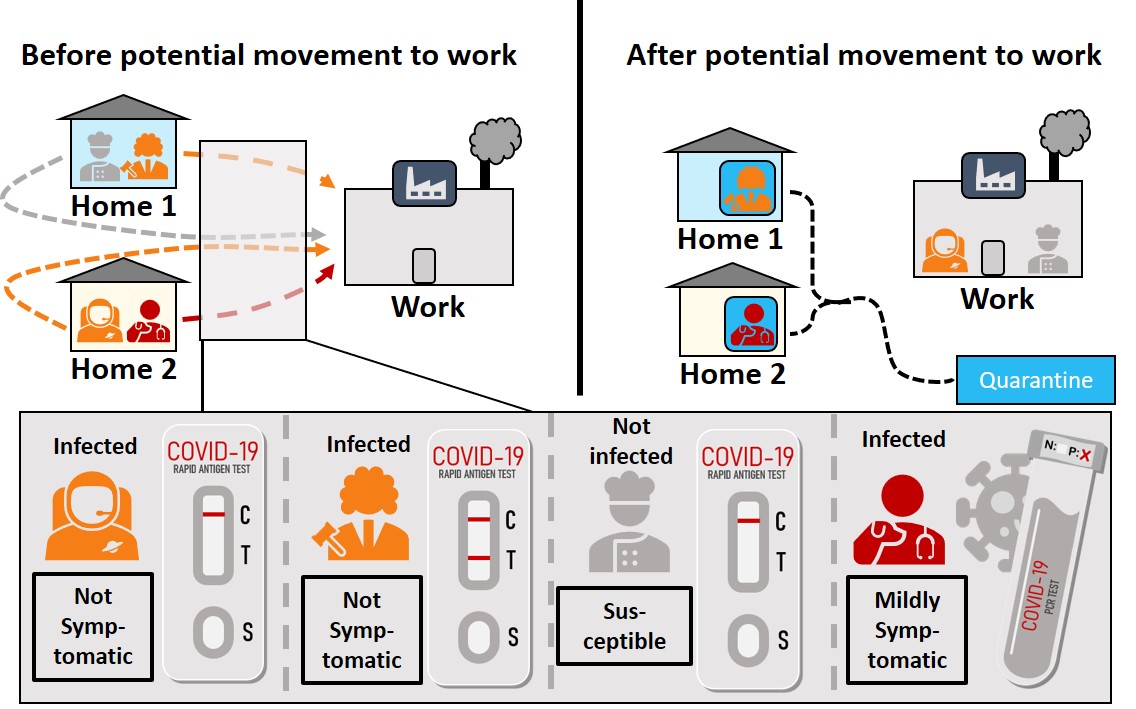} 
	\caption{\textbf{Example of Testing Schemes 1 and 2 from~\cref{fig:test_strategies}.} 
 In Testing Scheme 1, agents are obliged to test themselves before going to work. It is sufficient to use a Rapid-Antigen Test if they do not show symptoms. All agents who are not symptomatic, i.e., the first three agents in the bottom row, do an antigen test from which the first is false-negative, the second positive and the third negative. Through Testing Scheme 2, a PCR-Test is required if the agent shows symptoms. The last remaining, symptomatic agent in the bottom row performs a PCR test which turns out to be positive. The two positive tests resulted in quarantining at home.}
	\label{fig:test_strategy}
\end{figure}

Naturally, suitable testing and isolation can have a major impact on infectious disease dynamics by contributing to a reduction of secondary infections~\cite{salathe_covid-19_2020}. To allow for an advanced assessment of different testing strategies, every simulation can hold a \textit{Testing Strategy}. These strategies are composed of one or several \textit{Testing Scheme}s which are defined by different \textit{Testing Criteria}. A Testing Scheme comprises different Testing Criteria such as when, who, where (particular locations or all locations) and how. The "when" addresses the time period negative tests are required. The "who" addresses particular age groups or symptom states or it can also be untargeted, i.e., testing all agents. The "where" addresses the locations where tests are required to enter. The "how" addresses which type of test has to be used, how long a negative test from before is seen as valid and with which probability a test is done and checked for. Through the latter, we also implicitly implement incomplete control of testing restrictions. 
See~\cref{fig:test_strategies} for an example of a
Testing Strategy with three different Testing Schemes, 
and, in particular, see~\cref{fig:test_strategy} 
for the realization of Testing Scheme 1 and 2.

Different tests like PCR or Antigen-Rapid-Tests contain different parameters (sensitivity and specificity) and additional or other test types can be user defined. The user can specify a method to determine if a person is currently counted as infected, noting that studies reveal that the active virus can be traced shortly before symptom onset and usually lasts around 10 days after symptom onset for mildly symptomatic persons, while for severely symptomatic persons this may be longer~\cite{walsh_duration_2020,puhach_sars-cov-2_2023}.
In the base case, a person is considered to be infected, when it is not in either of the states Susceptible, Exposed, Recovered or Dead, and considered not infected otherwise.

If an agent wants to enter a location, where a test is required, i.e., a location-specific testing scheme that targets the agent, the test is performed (unless a negative test from the validity period is available to
the agent) and, based on the result, entry is allowed or denied. 

Although countywide rules for handling positive tests and resulting quarantine measures only became available in September 2021 through the "Niedersächsiche Absonderungsverordnung"~\cite{niedersachen_archiv}, isolation was already performed beforehand, in particular when getting contacted by the local health authorities. Thus, after positive testing, the agent is sent into quarantine. In the simulation, this leads to a movement to the home location of the agent, where it isolates itself.
In home locations, the agent's infectiousness is reduced to simulate distancing from other household members. 
If the reduction is not 100~\%, isolation is imperfect and secondary cases in households are possible.

Finally, agents also test voluntarily (with particular likelihoods) when going to a location, even if no mandatory testing scheme is in place. In the numerical section, we provided results for different probabilities of voluntary testing of symptomatic and nonsymptomatic individuals.

\subsubsection{Venue restrictions and closures}
Specific venue restrictions played a crucial role to stop the spread of COVID-19~\cite{brauner_inferring_2021}. The MEmilio ABM implementation
allows for several options. 

First, the model allows to completely close (particular) locations. Second, it is possible to adjust the capacity of locations, and third, to adjust the likelihood of particular agents to move to a location. By this, we can model a reduced percentage of persons to enter\SK{,} such as only 50~\% of pupils attending school on a particular day. If a capacity threshold is reached, further agents are not allowed to enter the location.
For every location, the particular contact matrix can be scaled to simulate contact reduction and social distancing.

\subsubsection{Masks and Compliance}

Masks are a crucial NPI to reduce the spread of the virus but their overall effectiveness is highly dependent on correct wearing and compliance~\cite{howard_evidence_2021}. Therefore, our model also implements the usage of different types of \textit{Mask}s such that locations can require agents to use a particular mask -- in this case stricter or more effective masks can also be worn to enter a location. 
In our simulations, a mask reduces the amount of virus particles emitted from an infectious agent and, on the other hand, it reduces the amount of virus particles received by other agents. For details, see~\eqref{eqn:aggregated_reduced_viral_shed} and~\eqref{eqn:infection_rate_indiv}. Through this implementation on the individual level, the effectiveness of mask wearing by highly vulnerable persons could be studied with the software.
We, furthermore, introduced individual compliance to mask wearing. With the maximum negative compliance value, a person never adheres to any mask mandate, while persons with maximum positive compliance wear a mask even if it is not mandatory. Furthermore, while compliance can be agent-specific, one agent can have different compliance values for wearing masks in different location types.

\subsection{Vaccination and immunity}

Vaccination has been the most critical pharmaceutical intervention during the SARS-CoV-2 pandemic, saving many lives~\cite{BOBROVITZ2023,Stein2023,bergstrom_counterfactual_2024}. In our model, vaccination provides agents with protection against severe illness:
When vaccinated, agents are less likely to develop severe symptoms, increasing their likelihood of transitioning from symptomatic to recovered.

To further capture immune response, we track each agent’s infection history, distinguishing between immunity gained from prior infections and different vaccines. This is reflected in varying protection factors depending on the type of immunity.

Since immunity wanes over time~\cite{BOBROVITZ2023,Stein2023}, our model incorporates this by using a piecewise linear function to gradually reduce an agent's protection factor. However, in this study of spring 2021, where vaccination just started to take up, we only model a constant protection factor $p_f$ against severe immunity.

\subsection{Stochasticity and random number generator}\label{sec:rng}
\SKSR{Stochasticity plays a major role in accurately modeling epidemic dynamics, as infection transmission and disease progression inherently involve random processes. Therefore, it must be carefully integrated to ensure realistic simulations.}
In order to \SK{model} stochastic events in the MEmilio ABM, we use a pseudo random number generator (PRNG). From a technical side, the choice of PRNG is important and the PRNG implementation has to satisfy several constraints. The PRNG has to be of high quality, lightweight, and lead to reproducible results for sequential and parallelized runs. A single global PRNG is inefficient in a multithreaded program because of the synchronization required. Thus, many parallel PRNGs are required that are efficiently initialized and that together generate a high\SK{-}quality sequence of random samples. If every agent has its own PRNG, the result of a parallelized simulation is deterministic even if the order of computation changes, e.g., due to scheduling or different thread counts. Counter-based PRNGs (cbPRNG) fulfill these requirements very well as described by~\cite{salmon_parallelrng_2011}.

In a cbPRNG, the state that gets updated every time a sample is generated is just a counter that is incremented. The actual sample is generated by a hash function $ h(k, c) $, where $ c $ is the counter and $ k $ is a key. The key can be seeded with real entropy like the state of a regular PRNG or set to a fixed value to reproduce the same sequence as in a previous simulation. There is a variety of suitable hash functions that are highly optimized for modern computer architectures, allowing fast generation of samples.

For a counter of size $B$ bits, the random sequence of $N = 2^B$ samples can be split into $L$ subsequences of $b$ bits with $ n = 2^b = N / L $ samples by starting the counter for the $i$-th subsequence at $i \cdot n$ for $i = 0,...,L-1 $. Thus, the cbPRNG is trivially parallelizable. We assign each agent one subsequence. For each agent, we store only the counter $\widetilde{c}_{i}$ in its own local subsequence. To get the corresponding counter $c_{i}$ in the global sequence from the local counter, we combine the local counter with the index of the agent as $ c_{i} = \widetilde{c}_{i} + i \cdot n $, efficiently implementable using bitwise operations. The subsequence index $i$ could be determined from the storage and iteration order of the agents. But in this model, we already store this index for other reasons. No additional storage is required either way. The key is stored only once. Only $b$ bits 
of storage per agent are required as opposed to regular PRNGs where each agent would need to store the full state, which is at least the size of the samples generated, more for high\SK{-}quality generators.

We are using Threefry-2$\times$32 implemented in the Random123 library~\cite{random123} and described in~\cite{salmon_parallelrng_2011}. This cbPRNG uses 64\SK{-}bit keys and counters and efficiently generates 64\SK{-}bit samples of sufficient statistical quality. Currently, we split the sequence into $2^{32}$ subsequences (i.e., up to roughly 4.3 billion agents) of $2^{32}$ samples. These numbers can be tuned in the future to support more agents, optimize storage, generate longer sequences, or increase the statistical quality further.

\subsection{Parallelization}\label{sec:parallel}
\SKSR{As agent-based models typically scale with the number of agents, an efficient implementation is essential to ensure computational feasibility. Furthermore, we can make use of parallelization to significantly reduce the simulation runtime, enabling the analysis of larger and more complex scenarios.}
We optimized our ABM simulation using parallelization in two different ways by employing both shared and distributed memory approaches. 
First, in order to optimize a single simulation run we utilized multicore shared-memory parallelism on single node machines, e.g., consumer laptops.
Second, we parallelized simultaneous independent runs, e.g., for calibration and uncertainty analysis on several cluster nodes using multicore as well as multinode parallelism combining both OpenMP and MPI.

For a single simulation, we parallelize both \textit{for} loops in~\cref{code:simulation_1} (interaction and movement) with OpenMP.
As we cache the exposure rate from each agent for each location before the interaction, changes of one agent 
do not 
affect the other agents during one time step. 
The parallel iteration in the interaction part does not require any synchronization. 
The movement iteration currently requires synchronization, since each location stores a list of agents at the location. These lists need to be protected by one mutex per list. Being not the focus of this paper, future work will consider removal of the mutex through changes to the architecture to improve scalability further.

For multiple independent concurrent simulations, we run these in parallel either on a single node using OpenMP or over several nodes using MPI in combination with OpenMP.
As we use nested parallelism within OpenMP, we specify first how many simulations to run in parallel on one node. The number of parallel simulations per node may be bounded by the available memory. The remaining cores can be used to parallelize single simulations, as described above.
Finally, when using multiple nodes, we distribute the simulation runs across nodes using MPI. On each node, simulations are run in parallel using OpenMP as described above. Initial setup and final data gathering requires communication between nodes. Simulations themselves are independent of each other and do not require any communication. Thus\SK{,} we are able to complete several thousand runs for the calibration and for uncertainty analysis in a matter of hours, e.g., we ran $85\,536$ simulation runs parallelized on 27 nodes (3\,456 cores) of JURECA-DC~\cite{julich_supercomputing_centre_jureca_2021} in less than eight hours.

\section{Results}\label{sec:Results}

In this section, we present various results for our ABM. We subdivide the section into five subsections.
\begin{itemize}
    \item In \cref{sec:perf}, we provide sequential performance and parallel scaling result.
    \item In \cref{sec:CaseStudy}, we conduct an extensive parameter fitting and validation process for the city of Brunswick between March 1st and May 31, 2021.
    \item In \cref{sec:Variation}, we present aggregated joint effects for different testing and isolation parameters.
    \item In \cref{sec:HighTesting}, we present time-resolved results for different symptomatic testing rates.
    \item In \cref{AlternateReality}, we consider a counterfactual scenario, where the lockdown has been replaced by substantially increased testing frequencies.
\end{itemize}

The memory usage of~\cref{sec:perf} has been measured on a Mac with an M1 Processor (4 Performance and 4 Efficiency Cores) and 8 GB of RAM. All timing runs have been conducted on JURECA-DC supercomputing facility~\cite{julich_supercomputing_centre_jureca_2021} equipped with compute nodes of 2x AMD EPYC 7742, 2× 64 cores, totaling to 98,304 CPU cores and a minimum of 512 GB DDR4 RAM, 3200 MHz, per node. 

\subsection{Performance and scaling}\label{sec:perf}

Before presenting the results of our ABM applied to a particular COVID-19 setting, we present performance and scalability results. We first provide performance results for a fixed number of cores: We use eight cores with a shared memory OpenMP parallelization 
(of eight threads) to also show the applicability of our ABM on any modern laptop computer. 

% population_size = [25, 50, 100, 200, 400, 800, 1600, 3200, 6400]  # Population size in thousand
% cpu_time_real = np.array([0.14, 0.36, 0.9, 1.8, 3.9, 7.92, 15.61, 31.2, 65.5])*(1/120.0)  # CPU time in seconds

% runtime = [ [0.9, 1.3, 1.5, 1.9, 2.6, 3.9],  0.025 mio agents per processor
%             [2.5, 2.8, 3.0, 3.9, 5.4, 7.0],  0.05 mio agents per processor
%             [5.2, 5.8, 6.3, 7.9, 10.4, 14.7],   0.1 mio agents per processor
%             [10.6, 11.1, 12.8, 15.6, 21.1, 32.1]]    0.2 mio agents per processor
% processors = [1, 2, 4, 8, 16, 32]  # Number of processors

\begin{figure}
	\centering 
    \includegraphics[width=0.48\linewidth]{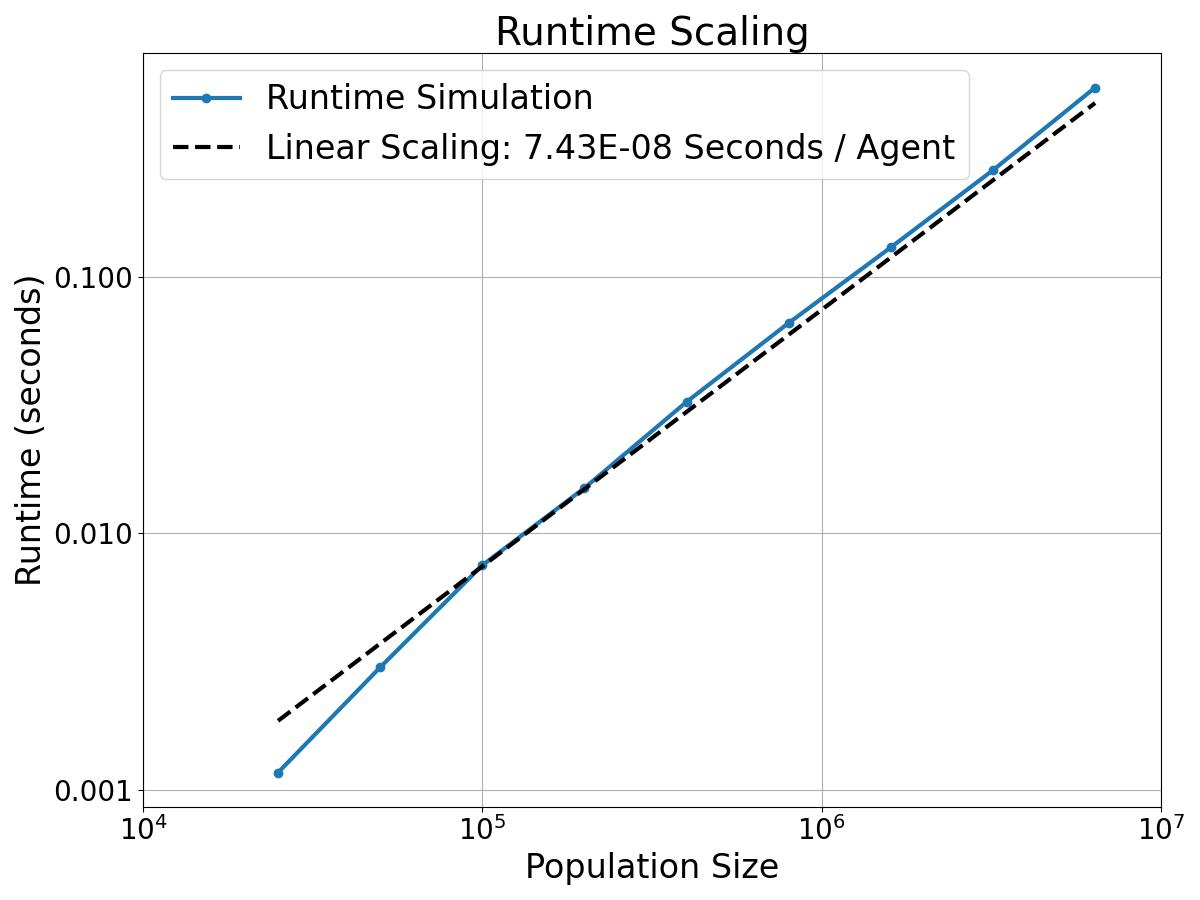} 
\hspace*{0.25cm}
	\includegraphics[width=0.48\linewidth]{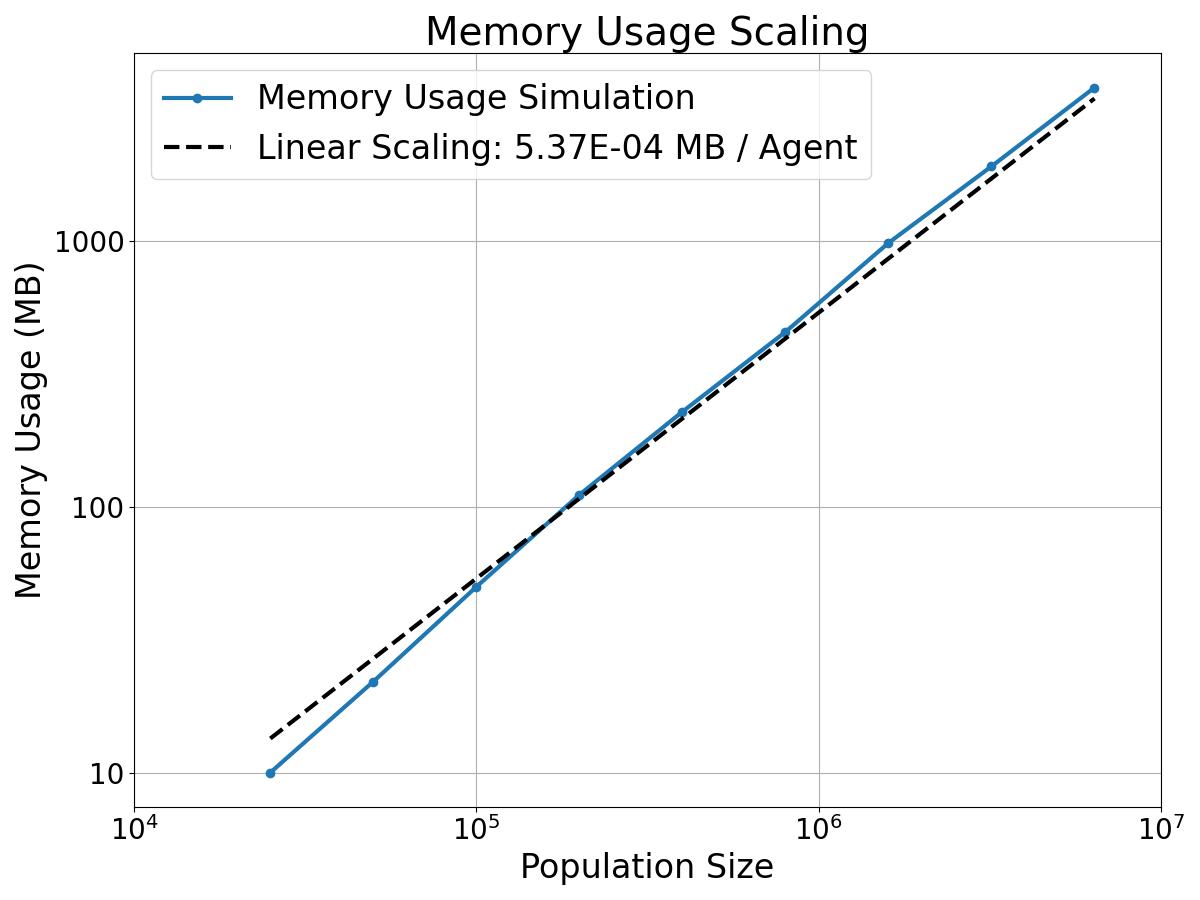} 
	\caption{\textbf{Runtime per time step (left) and (minimal) memory usage (right) of the MEmilio ABM against simulated number of agents.} Runtime and memory usage are shown in blue and a dashed linear scaling line in black is provided as a reference line.}
	\label{fig:linear_time_mem_scaling}
\end{figure}

\begin{figure} 
	\centering \includegraphics[width=0.9\linewidth]{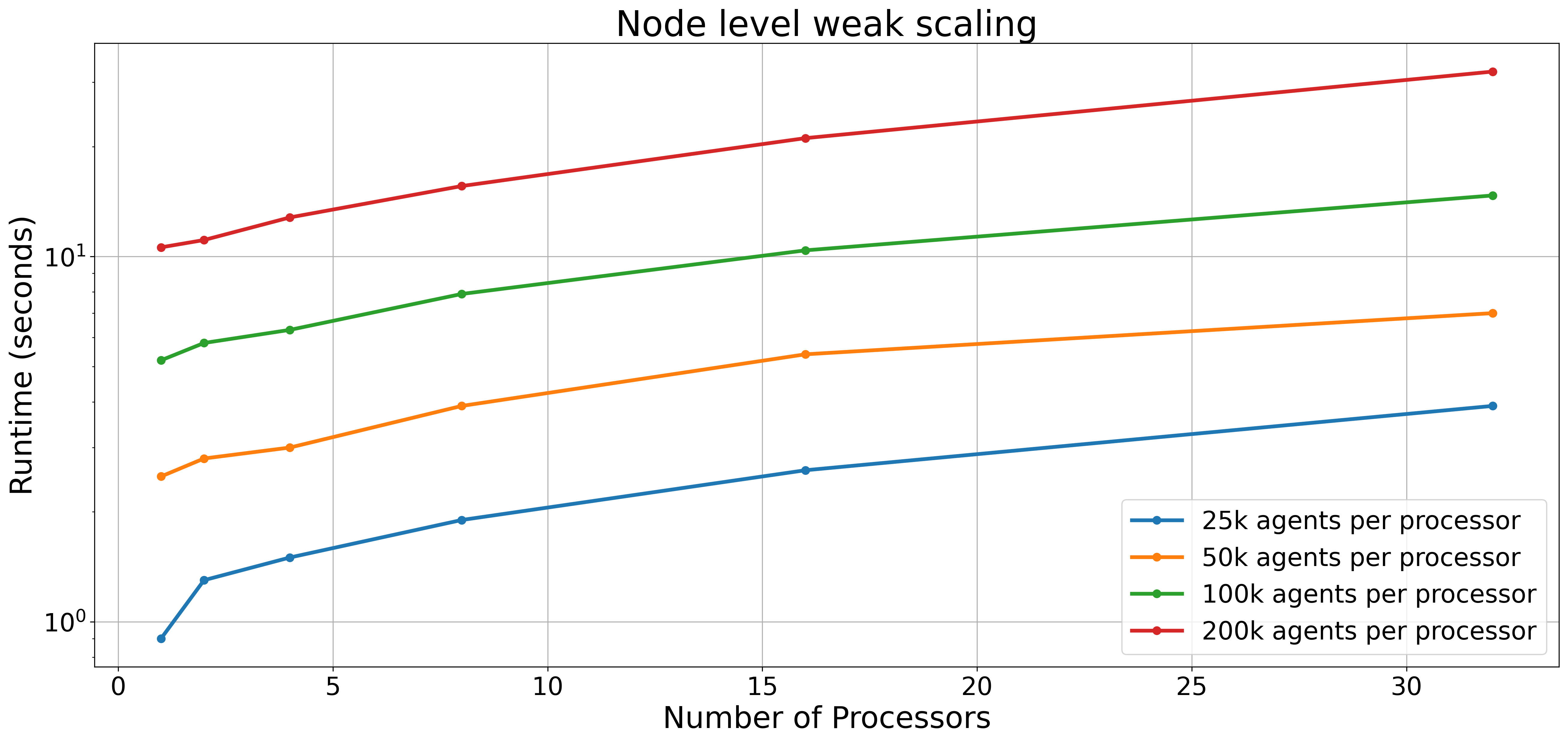} 
	\caption{\textbf{Node-level weak scaling of the MEmilio ABM with different numbers of agents per processor}. Runtime per time step \SK{for up to 32 cores}.}
	\label{fig:weak_scaling}
\end{figure}

In~\cref{fig:linear_time_mem_scaling}~(left), we see that the runtime (given per time step) scales linearly with the number of agents. 
For stability of the timings, the runtime (per time step) has been computed through the average over a simulation of five days. 
The simulation of one million agents takes 0.08 seconds per time step such that a one-day-simulation (24 time steps) takes roughly two seconds. This 
allows a simulation of one million agents over one month in roughly one minute on a laptop computer. 
Correspondingly, the simulation of 100\,000 agents over a full year takes roughly one minute.

In~\cref{fig:linear_time_mem_scaling}~(right), we see that the memory usage (with minimal logging for I/O) also scales linearly with the number of agents and that we need roughly 50 MB for one million agents. With user-defined loggers, additional logging and output over the simulation can be controlled for most of the variables so that the memory usage can be kept to a minimum of what is needed for the user's analysis.

In~\cref{fig:weak_scaling}, we provide node-level weak scaling results for our ABM with different numbers of agents per processor, with detailed timings in~\cref{tab:weak_scaling}. 
\SK{While the efficiency scales well up to four cores, it declines significantly between 8 and 32 cores, however, still achieves $33~\%$ efficiency for a 32-fold increase in problem size.}
Future research will consider memory saturation issues, optimizing load balancing, and improving synchronization.

\begin{table}
\centering
\caption{\textbf{Node-level weak scaling, 200k agents per core.} Runtime per time step \SK{for up to 32 cores}.}\label{tab:weak_scaling}
\begin{tabular}{c|c|c|c}
\#cores & total agents & runtime(s) & efficiency \\
\hline
\hline
1   & 200\,000  & $8.83e-02$ & 100~\% \\
2   & 400\,000  & $9.25e-02$ & 95~\% \\
4   & 800\,000  & $1.07e-01$ & 82~\% \\
8   & 1\,600\,000  & $1.30e-01$ & 66~\% \\
16   & 3\,200\,000  & $1.76e-01$ & 50~\% \\
32   & 6\,400\,000  & $2.68e-01$ & 33~\% \\
\end{tabular}
\end{table}

\subsection{Fitting and validation on a Brunswick scenario}\label{sec:CaseStudy}

\subsubsection{Model setup and parameter overview}\label{sec:Parameters}

\begin{table}[]
\begin{adjustbox}{width=\textwidth}
\begin{tabular}{l l l l}
\toprule
\textbf{Variable} & \textbf{Description} & \textbf{Value(s) or Distribution} & \textbf{Source and explanation} \\
\midrule
$v_{\textrm P}$ & Viral load peak & $8.1$ & Motivated by ~\cite{jones_estimating_2021} \\
$v_{\textrm I}$ & Viral load incline & $2$ & Motivated by ~\cite{jones_estimating_2021} \\
$v_{\textrm D}$ & Viral load decline & $-0.17$ & Motivated by ~\cite{jones_estimating_2021} \\
$\alpha$ & Viral shed parameter & $-7$ & Motivated by ~\cite{jones_estimating_2021} \\
$\beta$  & Viral shed parameter & $1$ & Motivated by ~\cite{jones_estimating_2021} \\
$s_{\textrm f}$    & Viral shed factor & Gamma($1.6, 1/22$) & Motivated by ~\cite[Fig. 3]{ke_daily_2022} \\
$\lambda$    & Infection rate from viral shed & $1.596$ & Grid search \\
\midrule
$p_{\textrm {Sym}}$       & Chance to develop symptoms from an infection & $\{0.5, 0.55, 0.6, 0.7, 0.83, 0.9\}$ & ~\cite[Tab. 2]{kerr_covasim_2021} \\
$p_{\textrm {Sev}}$       & Chance to develop severe symptoms from a symptomatic infection & $\{0.02, 0.03, 0.04, 0.07, 0.17, 0.24\}$ & ~\cite{nyberg_risk_2021} \\
$p_{\textrm C}$           & Chance to develop critical symptoms from a severe infection & $\{0.1, 0.11, 0.12, 0.14, 0.33, 0.62\}$ & ~\cite[Tab. 2]{zali_mortality_2022} \\
$p_{\textrm D}$           & Chance to die from a critical infection & $\{0.12, 0.13, 0.15, 0.26, 0.4, 0.48\}$ & ~\cite[Tab. 2]{zali_mortality_2022} \\
$t_{\textrm E}^{\textrm N}$         & Time from Exposed to nonsymptomatic & LogNormal($4.5, 1.5$) & ~\cite[Tab. 1]{kerr_covasim_2021} and references within \\
$t_{\textrm N}^{\textrm {Sym}}$     & Time to develop symptoms after infection in case of a symptomatic infection & LogNormal($1.1, 0.9$) & ~\cite[Tab. 1]{kerr_covasim_2021} and references within \\
$t_{\textrm N}^{\textrm R}$         & Time to recover in case of an asymptomatic infection & LogNormal($8.0, 2.0$) & ~\cite[Tab. 1]{kerr_covasim_2021} and references within \\
$t_{\textrm {Sym}}^{\textrm {Sev}}$ & Time to develop severe symptoms in case of a severe infection & LogNormal($6.6, 4.9$) & ~\cite[Tab. 1]{kerr_covasim_2021} and references within \\
$t_{\textrm {Sym}}^{\textrm R}$     & Time to recover in case of a symptomatic infection & LogNormal($8.0, 2.0$) & ~\cite[Tab. 1]{kerr_covasim_2021} and references within \\
$t_{\textrm {Sev}}^{\textrm C}$     & Time to develop critical symptoms in case of a critical infection & LogNormal($1.5, 2.0$) & ~\cite[Tab. 1]{kerr_covasim_2021} and references within \\
$t_{\textrm {Sev}}^{\textrm R}$     & Time to recover in case of a severe infection & LogNormal($18.1, 6.3$) & ~\cite[Tab. 1]{kerr_covasim_2021} and references within \\
$t_{\textrm C}^{\textrm D}$         & Time to die in case of death & LogNormal($10.7, 4.8$) & ~\cite[Tab. 1]{kerr_covasim_2021} and references within \\
$t_{\textrm C}^{\textrm R}$         & Time to recover in case of a critical infection & LogNormal($18.1, 6.3$) & ~\cite[Tab. 1]{kerr_covasim_2021} and references within \\
\midrule
$m_{\textrm r}$ & Mask protection factor for the receiving person & $0.25$ & Motivated by ~\cite{kuhn_assessment_2021,koslow_appropriate_2022} and manual adaptation \\
$m_{\textrm t}$ & Mask protection factor for the transmitting person & $0.25$ & Motivated by ~\cite{kuhn_assessment_2021,koslow_appropriate_2022} and manual adaptation \\
$m_{\textrm c}$ & Overall mask compliance per location type & $\{0, -0.1, -0.1, -0.3, -0.2\}$ & Manual adaptation \\
$p_{\textrm f}$ & Protection from vaccination against severe infection & $0.8$ & Motivated by ~\cite{zunker_novel_2024} and manual adaptation \\
$q_{\textrm e}$ & Quarantine efficiency & $0.5$ & Manual assumption \\
$q_{\textrm d}$ & Quarantine length & $10$ & Manual assumption \\
\midrule
& Test Sensitivity & $0.71$ & ~\cite{fernandez-montero_validation_2021} \\
& Test Specificity & $0.996$ & ~\cite{fernandez-montero_validation_2021} \\
$p_{\textrm s}$ & Testing probability for symptomatic agents & $0.02472$ & Grid search \\
$\mu_{\textrm {ns}}$ & Testing ratio symptomatic to nonsymptomatic agents & $4.83$ & Grid search \\
& Relative increased testing ratio during lockdown & $0.2$ & Motivated by \cite{braunschweig_krisenstab_maerz} and manual adaption\\
\midrule
$d$ & Initial dark figure & $4.171$ & Grid search \\
$r_{\textrm L}$ & Relative contact reduction during lockdown & $0.2725$ & Grid search \\
& Reduced trips during lockdown (Basic Shop, Social Event) & $0.5$ & Motivated by ~\cite{kuhn_assessment_2021,koslow_appropriate_2022} and manual adaptation\\
& Reduced trips factor during lockdown (Work) & $0.7$ & Motivated by ~\cite{kuhn_assessment_2021,koslow_appropriate_2022} and manual adaptation \\
& Reduced trips factor during lockdown (School) & $0.0$ & Schools closed\\
 & Relative contact reduction after lockdown & $0.5$ & Manual adaptation due to reduction in positive tests. \\
& Reduced trips factor after lockdown (Basic Shop, Social Event) & $0.8$ & Motivated by ~\cite{kuhn_assessment_2021,koslow_appropriate_2022} and manual adaptation \\
& Reduced trips factor after lockdown (Work) & $0.75$ & Motivated by ~\cite{kuhn_assessment_2021,koslow_appropriate_2022} and manual adaptation \\
& Reduced trips factor after lockdown (School) & $0.5$ & Motivated by alternating lessons\\
$\psi$ & Seasonality factor April & $0.95$ & Motivated by ~\cite{kuhn_assessment_2021,koslow_appropriate_2022} and manual adaptation \\
$\psi$ & Seasonality factor May & $0.85$ & Motivated by ~\cite{kuhn_assessment_2021,koslow_appropriate_2022} and manual adaptation \\
$r_{\textrm E} $& Percentage of the population celebrating Easter & $0.2$ & Motivated by ~\cite{statista_ostern} and manual adaptation \\
\bottomrule
\end{tabular}
\end{adjustbox}
\caption{\textbf{Final ABM parameters used after the fitting process.} For the first four parameters, values are given for the different age groups in ascending order. The values for mask compliance are given for Home, School, Work, Social Event, and Basic Shop. \SK{Parameters not available in the literature are either estimated by a grid search based on real-world data, where feasible, or manually adjusted, potentially informed by relevant literature.} }
\label{tab:parameters}
\end{table}

For fitting and validation, we consider the period of March 1st to May 30th, 2021 for the city of Brunswick. 
The full set of parameters is shown in~\cref{tab:parameters}. While many parameters could be taken from literature, some key quantities were not known and were fitted by us. We use the age groups 0 to 4, 5 to 14, 15 to 34, 35 to 59, 60 to 79, and 80+ life years as these are also used in most of the reported data for Germany~\cite{rki_cases}.

Most virus specific parameters are available through studies and literature:
All viral load parameters are taken from~\cite{jones_estimating_2021}, except for the distribution of individual viral sheds, $s_f$, which is estimated from~\cite[Fig. 3]{ke_daily_2022}.

We use the LogNormal-distributed infection state times in infection states $t_{\textrm E}^{\textrm N}$, $t_{\textrm N}^{\textrm {Sym}}$, $t_{\textrm N}^{\textrm R}$, $t_{\textrm {Sym}}^{\textrm {Sev}}$, $t_{\textrm {Sym}}^{\textrm R}$, $t_{\textrm {Sev}}^{\textrm C}$, $t_{\textrm {Sev}}^{\textrm R}$, $t_{\textrm C}^{\textrm D}$ and $t_{\textrm C}^{\textrm R}$ that were also used in~\cite[Tab. 1]{kerr_covasim_2021}.
For the probability to develop symptoms, $p_{\textrm {Sym}}$, we similarly refer to~\cite[Tab. 2]{kerr_covasim_2021}.

For the probability to develop severe symptoms, $p_{\textrm {Sev}}$, which requires hospitalization, we use the cohort analysis~\cite{nyberg_risk_2021} adapted to our age groups.

The probability to die, $p_D$, is determined with the help of the multi-center study~\cite[Tab. 2]{zali_mortality_2022}, where a distinction is being made between death rates of non-ICU patients and ICU patients. However, in our model, critically infected agents will always receive ICU treatment, and only these agents may die. This is not reflecting agents dying before receiving ICU treatment or with the lack of available ICU spots, which would still be listed as severely infected agents. To account for this fact, we use the death rate for ICU patients only in our model and adjust the probability to develop critical symptoms, $p_C$, accordingly, such that the total death rate is recovered.
In the shown graphs, this artificial increase in critically infected agents is reversed, so that the correct amount of both severely infected agents and critically infected agents is recovered.

For the mask protection, we only use one generic mask type as we could not identify the percentage of medical or FFP2 masks. We started the iteration from an aggregated effect of 25-35~\% as used in a macroscopic model ~\cite{kuhn_assessment_2021,koslow_appropriate_2022}. We finally set emitting and receiving reduction by 25~\%, resulting in a (squared) transmission reduction by roughly 44~\%.

Further, the number of conducted self or rapid tests has not been reported. As we suppose their amount to be largely superior to the amount of conducted PCR tests and as PCR test were often used to confirm a positive self test, we only use (antigen) tests with a sensitivity of $0.71$ and specificity of $0.996$; cf.~\cite{fernandez-montero_validation_2021}. As the modeled specificity is high, we do not assume substantial deviations through false positive agents.

Quarantine has been discussed in~\cref{sec:testing}. We assume a base quarantine length of $10$ days and a quarantine efficiency of $0.5$, and investigate parameter variations in quarantine length and efficiency in~\cref{sec:Variation}.

During the considered time period, we model seasonality with factors of 0.85 and 0.95 in April and May. For these, we also started from factors as validated in~\cite{bauer_relaxing_2021,kuhn_assessment_2021} and manually adapted. 
While the seasonal factors are applied on the viral spread, we use an additional contact reduction factor
of 0.5 after lifting of the lockdown to model persisting cautious behavior and more outdoor activities.

\subsubsection{Model initialization}\label{sec:Model_init}

We use the data set of~\cite{schengen_2024_13318436} with 373\,378 agents\DKSR{. This includes the base population that lives in Brunswick plus the incommuters and their households.} We use the officially reported case data from~\cite{rki_cases} as well as reported ICU admissions~\cite{rki_divi}.
As the officially reported case data does not report on an individual basis, but aggregates on local geographic and age group levels, and as death numbers are added only retrospectively to the case numbers, exact death days are not known. Therefore, we extrapolate the day of death from the day of infection and our model parameters. We assume the reporting date $t_{\textrm {rep}}$ of an infection to be close after symptom onset and thus estimate time of death as
\begin{equation}
    t_{\textrm D} = t_{\textrm {rep}} + \lfloor t_{\textrm {Sym}}^{\textrm {Sev}} + t_{\textrm {Sev}}^{\textrm C} + t_{\textrm C}^{\textrm D} \rfloor,
\end{equation}
where we use the mean values for the time in infection states $t_{\textrm {Sym}}^{\textrm {Sev}}$, $t_{\textrm {Sev}}^{\textrm C}$, and $t_{\textrm C}^{\textrm D}$ from~\cref{tab:parameters},  rounding down to the next whole day. For our parameters, this leads to an estimated time of $18$ days between initial report and death. The infected population is then initialized from the product of the officially reported data and the fitted dark figure (see below).

Vaccinations started to roll out at the beginning of 2021 and affected mainly older people. We use real-world infection data of Brunswick~\cite{rki_vacc} to vaccinate persons accordingly in silico.

Within the three month period, 
the city of Brunswick went into a lockdown when its incidence went above 100~\cite{rki_cases}. Before and after the lockdown, several lighter interventions like alternating lessons or remote work recommendations were in place. With the lockdown, stricter interventions such as school closures and contact restrictions had been implemented. These measures have been adapted to our model and are used in the lockdown scenarios. The lockdown spans from March 30th to April 29th, 2021~\cite{braunschweig_krisenstab_maerz,braunschweig_krisenstab_april}. Over the whole time, we have a maximum of 50~\% of pupils going to school, while in the lockdown period schools are completely closed. We model a reduction of 20~\% of persons going to shops and social events, increased to 50~\% reduction during the lockdown. We further assume a 25~\% reduction in work-related trips due to remote work, increased to 30~\% during the lockdown.

Furthermore, due to the special significance of Easter in Germany, we include Easter as a special event in our simulations. During Easter, the positive tests declined by roughly 34~\%, see~\autoref{fig:fitting4things}. We simulate that, in addition to the regularly performed trips, people meet at Easter Sunday or Monday to celebrate Easter in family gatherings. According to~\cite{statista_ostern}, approximately 34~\% of persons celebrated Easter with their family in 2019. Conservatively, we estimate the number to be reduced to $r_{\textrm E} = 20~\%$ in 2021 due to the pandemic and the active lockdown.

\subsubsection{Parameter fitting}\label{sec:Fitting}

\begin{figure}
    \centering
    \includegraphics[width=1.0\linewidth]{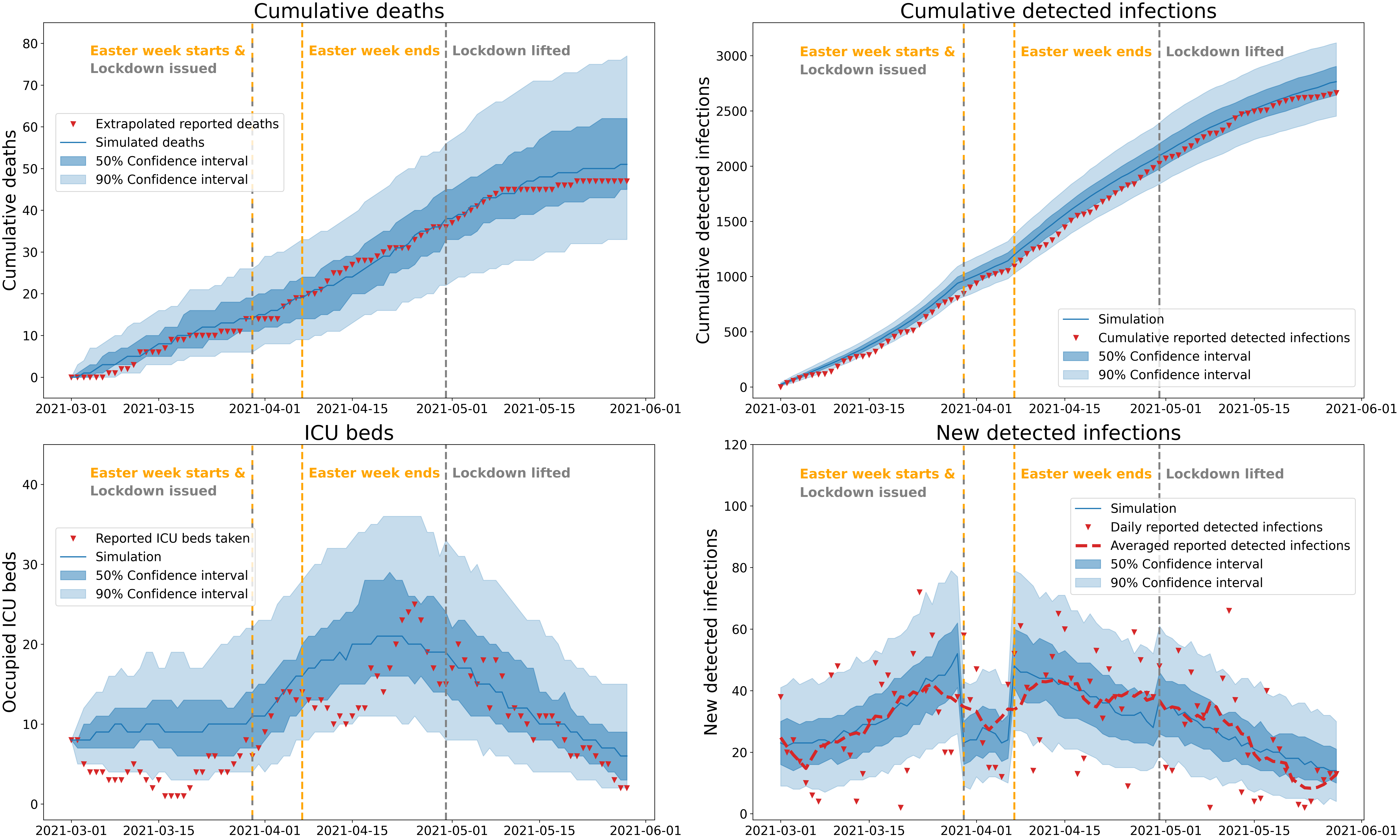}
    \caption{\textbf{Simulated outcomes of the fitted ABM for Brunswick between March 1st and May 30, 2021.} Median simulation results \SK{with the best-fit parameters} are provided with a solid blue line, percentiles p25 and p75 as well as p5 and p95 are shown through differently shaded, blue regions. Reported data shown in red. Cumulative deaths (top left), cumulative detected (top right), ICU occupancy (bottom left) and newly detected infections (bottom right) are shown as outcomes. \SK{Vertical lines indicate the beginning of the Easter week and start of the lockdown (left line), the end of the Easter week (middle line) and the end of the lockdown (right line).}}
    \label{fig:fitting4things}
\end{figure}

In order to fit the parameters to the data, we use an exhaustive grid search to minimize a weighted mean squared error (MSE). 
In this weighted MSE, we compare the reported versus simulated data for the cumulative number of COVID-19 related deaths, the persons treated in ICU on a daily basis as well as the cumulative positive tests over the whole 90 days of the simulation. 
The sum of each MSE is added by an 10000:1000:3 weight ratio, respectively, addressing the different magnitudes of the time series. This ratio leads to a reasonable fit to all three data sets and was chosen through an iterative process.
As the complete search space for even a small number of parameters in ABMs is often tremendously high, we focused on key parameters that could not be estimated well from the literature. 
We have identified five key parameters to fit, which are the linear infection coefficient $\lambda$, the initial dark figure $d$ which describes the ratio of all cases to the detected cases at the beginning of the simulation, the contact reduction when the lockdown happens $r_{\textrm L}$, the testing rate of symptomatic persons $p_{\textrm s}$ and the ratio of asymptomatic testing probability compared to symptomatic testing probability $\mu_{\textrm {ns}}$; see also~\cref{tab:parameters}.

In general, we run every parameter set 11 times 
to address the stochastic nature of our model and average the MSEs. For simplicity, we chose an odd number to have a direct representation of the median.
From these outcomes, the best average MSE is selected as the best fit. 
%mmh For comparison, in a similar setting, the authors of~\cite{robertson_bayesian_2024} ran 50 different seeds for each of the 700 parameter sets in a four-dimensional space to successfully fit a model for Chicago.

As we want to have an exhaustive exploration of the parameter space, we split the grid search into a two-step process.
First, we define broad intervals for each of the five parameters. Then, per dimension, we select six equally distributed parameter values in these intervals.
For five parameters this already results in $7\,776$ parameter combinations and thus to $85\,536$ simulation runs.
In a second step, we take the three best parameter combinations. For each of these parameter combinations, we run a second grid search spanning intervals around the previously found values, allowing a maximum deviation of 5~\% in both directions.
Again, this interval is divided into six points for each dimension and the best MSE is taken as the final fit.

After finalization of the fitting process, we conduct 128 simulations of the ABM and provide the results \SK{and parameters, where the lowest MSE was attained,} on cumulative deaths and detected infections as well as ICU occupancy and daily detected new infections in~\cref{fig:fitting4things}. \SK{Here we infer the new detected infections from the reported cumulative detected infections.}

The fitted parameters are shown in~\cref{tab:parameters}. We obtain an initial dark figure $d = 4.171$, which is in the range of plausible values; see, e.g.,~\cite{fiedler_mathematisches_2021} for 2020. Furthermore, we obtain a symptomatic testing probability of $p_{\textrm s} = 2.472~\%$ per individual trip, a ratio of ${\mu}_{\textrm {ns}} = 4.83$ for symptomatic against nonsymptomatic individuals to test. 

For the contact reduction factor during lockdown, we obtain a value of $r_{\textrm L} = 0.2725$. This value is smaller than prior assumptions that we have used in aggregated models in~\cite{kuhn_assessment_2021,kuhn_regional_2022,koslow_appropriate_2022}. However, in the ABM, it has been combined with the reduced number of trips for the particular locations school, work, shop, and social event and the increased protective values of masks. The fitted infection rate from viral shed is $\lambda = 1.596$.

\subsection{Effects from combined variation of testing and isolation parameters}\label{sec:Variation}

\begin{figure} 
	\centering \includegraphics[width=0.9\linewidth]{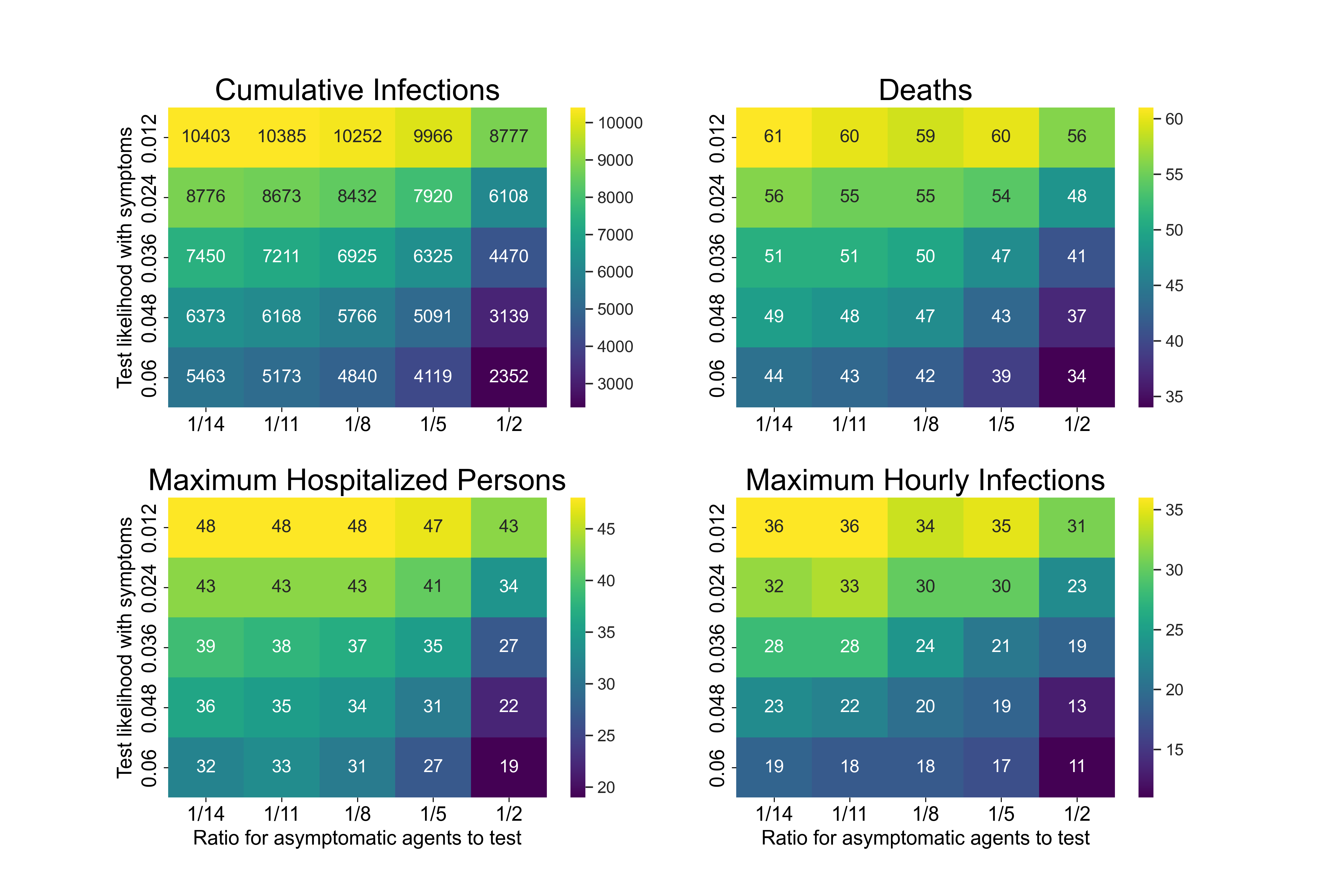} 
	\caption{\textbf{Aggregated outcomes for a variation of voluntary likelihood for symptomatic agents to test, $p_{\textrm s}$, against ratio for nonsymptomatic to symptomatic agents to test, $1/ \mu_{\textrm {ns}}$.} Aggregated outcomes on cumulative infections (top left), deaths (top right), maximum hospitalized patients (bottom left), and maximum \SK{hourly} infections (bottom right). Testing likelihood for symptomatic patients increases from top to bottom. Ratio for testing of symptomatic against symptomatic increases from left to right.}
	\label{fig:ParVarTest}
\end{figure}

Substantial changes in different parameters such as the likelihood for symptomatic testing naturally induce changed outcomes of our simulation. However, varying single parameters is often insufficient to get more insight into the model as combined effects are then overlooked. In this section, we want to explore the combined effects of different testing and isolation parameters. We present all combined effects with respect to four different outcomes: the cumulative number of infections, the number of deaths, the maximum number of hospitalized patients and the maximum number of \SK{hourly} infections. While it is clear why to consider the first two indicators, maximum numbers are of high importance as capacities of the health sector should not be attained.

First, we compare the likelihood of voluntary symptomatic testing, $p_{\textrm s}$, against the ratio of (voluntary) symptomatic against asymptomatic testing, $\mu_{\textrm {ns}}$. The voluntary likelihood to test is trip-based and for every trip conducted by an agent, the particular probability is used to determine whether an agent tests (see \cref{sec:testing}). 
From~\cref{fig:ParVarTest}, we see that weak symptomatic control (see top rows), i.e., a low likelihood to test symptomatic persons, implies that symptom-independent or untargeted nonsymptomatic testing strategies have limited or no effect to mitigate disease dynamics. That means, while this testing could still have brought down numbers and prevented additional infection surges, the resulting reduction factor is rather low, i.e., 8-16~\%, depending on the considered outcome. Furthermore, from top to bottom rows, we see that the better the symptomatic control, the more effective is the untargeted testing.

\begin{figure} 
	\centering \includegraphics[width=0.9\linewidth]{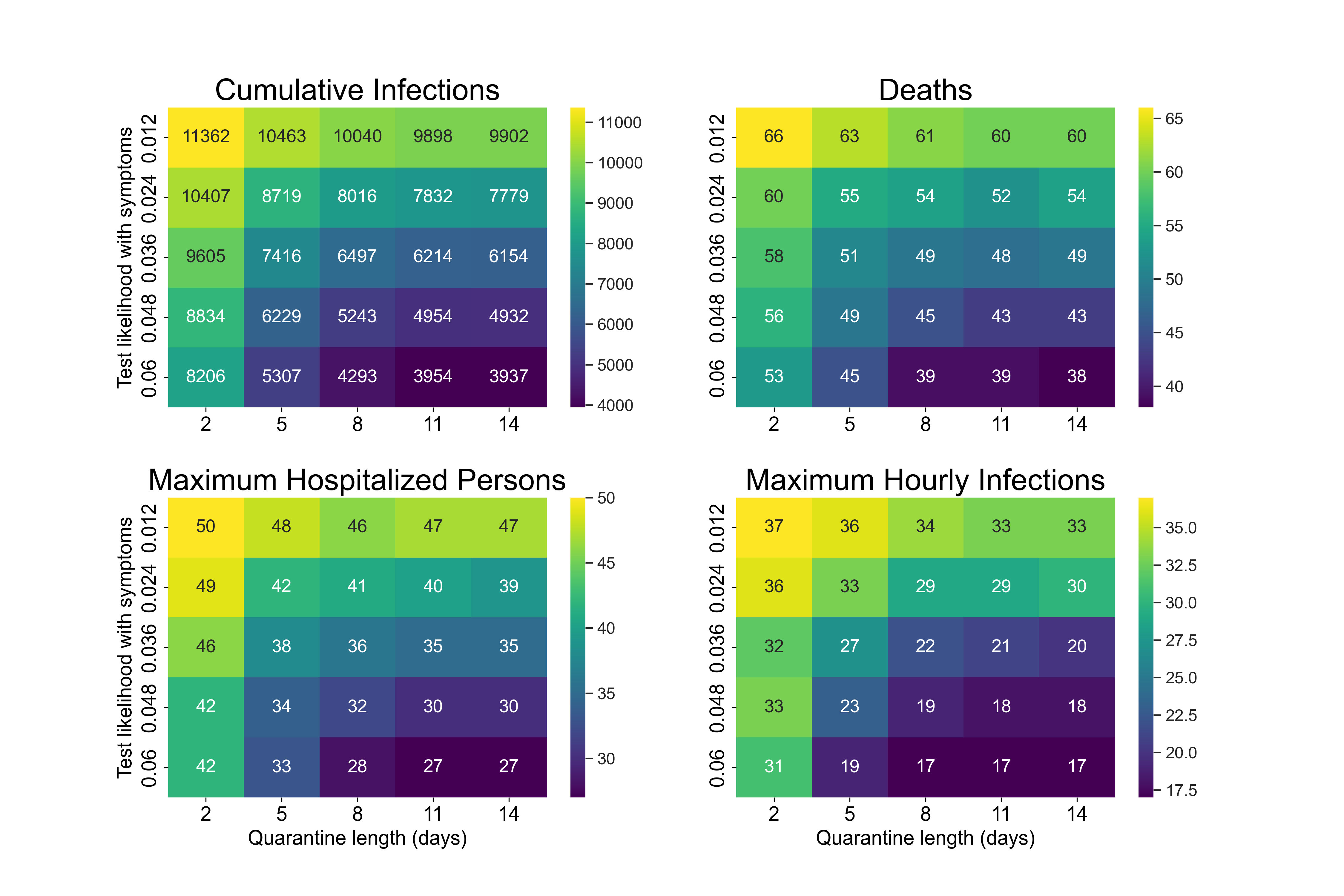} 
	\caption{\textbf{Aggregated outcomes for a variation of voluntary likelihood for symptomatic agents to test, $p_{\textrm s}$, against quarantine length, $q_{\textrm d}$.} Aggregated outcomes on cumulative infections (top left), deaths (top right), maximum hospitalized patients (bottom left), and maximum \SK{hourly} infections (bottom right). Testing likelihood for symptomatic patients increases from top to bottom. Quarantine length in number of days  increases from left to right.
 }
 	\label{fig:ParVarQuarTests}
\end{figure}

Second, we compare the likelihood of voluntary symptomatic testing against quarantine length. From~\cref{fig:ParVarQuarTests}, we see that the quarantine is the more effective, the better the symptomatic control. For the combination of these two parameters, we see that the expected single effects (more testing, better mitigation and longer quarantine, better mitigation) are also reflected in the combined effects view, with improvements in one dimension leveraging the effect in the other dimension.

\begin{figure} 
	\centering \includegraphics[width=0.9\linewidth]{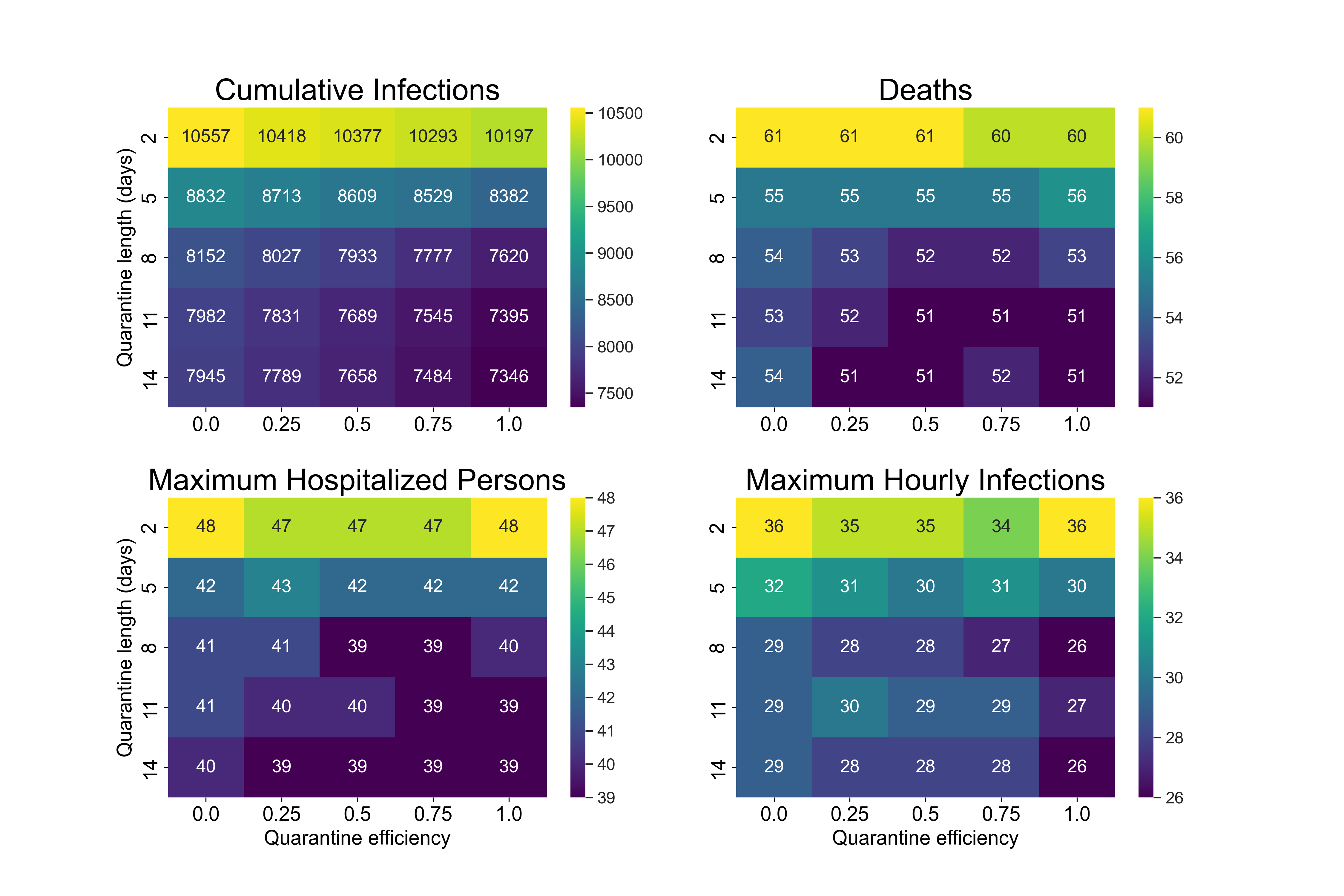} 
	\caption{\textbf{Aggregated outcomes for a variation of quarantine length, $q_{\textrm d}$, against quarantine efficiency, $q_{\textrm e}$.} Aggregated outcomes on cumulative infections (top left), deaths (top right), maximum hospitalized patients (bottom left), and maximum \SK{hourly} infections (bottom right). Quarantine length in number of days increases from top to bottom. Quarantine efficiency $q_e$ increases from left to right.}
	\label{fig:ParVarQuarantine}
\end{figure}

Third, we compare the quarantine length against the quarantine efficiency. Quarantine efficiency (or compliance) is often far from the perfect case of 1.0~\cite{zou_trade-off_2023} (and the references therein). From~\cref{fig:ParVarQuarantine}, we see that quarantine efficiency is almost irrelevant if the quarantine length is chosen too short (i.e., two days). In our simulations, most first positive tests (resulting in quarantine) are found to be performed shortly after symptom onset, which is close to peak viral load, as explained in~\cref{sec:transmission}. Thus, the time of quarantining usually overlaps well with the peak viral load and the slow decreasing phase after, which are most relevant to transmission. As expected, it is most important to quarantine in the beginning when viral load is high and the most substantial reduction in infections, hospitalizations and deaths is achieved with a quarantine length of 5–11 days and that there is little to no effect with longer quarantining. 
We can also conclude that, in our model, quarantine length is more important than quarantine efficiency. Combined with the results of the previous section, we can conclude that a higher testing rate for symptomatic persons could even make short quarantines have a substantial effect on mitigation. 
However, these findings are in the light of a fitted symptom-based testing rate as of March to May 2021.

\subsection{Increased symptomatic testing}\label{sec:HighTesting}

\begin{figure} 
	\centering \includegraphics[width=1.0\linewidth]{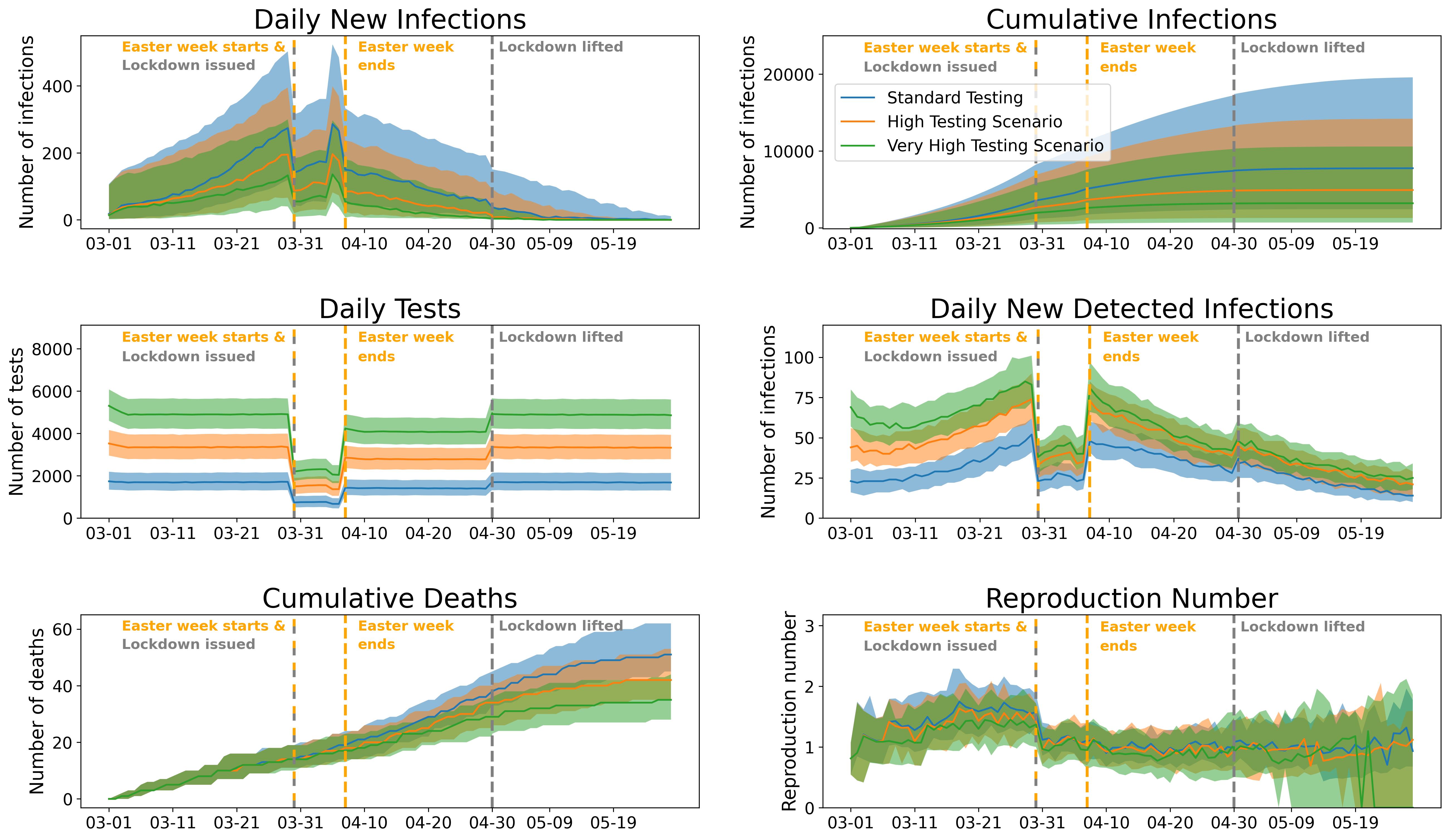} 
	\caption{\textbf{Impact of increased symptomatic testing.} In our ABM fitted to data from Brunswick between March 1st and May 30, 2021 (blue),
  the symptom-based testing probability was doubled (orange) and tripled (green), \SK{where the shaded area represents the 50\% confidence interval}. The lockdown was active from end of March until end of April, while test frequency was further reduced in the Easter week, \SK{ as indicated by vertical lines. Furthermore, there are special Easter gatherings on Easter Sunday and Monday.} }
	\label{fig:HighTesting}
\end{figure}

 While in the previous section we have considered the aggregated outcome of combined effects for symptom-dependent and -independent testing strategies, here we will present the detailed time series outcome for increased symptomatic testing. In~\cref{fig:HighTesting}, the blue curve represents the outcome for our fitted default likelihood to test on symptoms. With orange and green, we present the outcomes for a doubled and tripled likelihood of symptomatic individuals to test before a trip is started.

In order to estimate the currently effective reproduction number, we use an adaptation of the formulation of the instantaneous reproduction number from~\cite{cori_new_2013}:
\begin{equation}
    R_{\textrm t} = \frac{I_{\Delta {\textrm t}}}{\sum\limits_{p\in\Omega\,:\,s_{\textrm p}(t-\Delta t)>0\,\lor\,s_{\textrm p}(t)>0} \frac{\int_{t-\Delta t}^t s_{\textrm p}(\tau) \mathrm{d}\tau}{\int_{-\infty}^\infty s_{\textrm p}(\tau) \mathrm{d}\tau}}
\end{equation}
Here, $I_{\Delta {\textrm t}}$ is the incidence from the recent time period $[t-\Delta t, t)$.

While other computations of the reproduction number, such as the cohort reproduction number, are done retrospectively, knowing the actual amount of total secondary cases of every index case within the simulation framework, we opted to use this prospective formulation, as it provides a real-time estimation and can be used to act as an indicator for NPIs in the future development. However, using this formula in the following results, we suspect a slight overestimation of the actual reproduction number.
 
We see that the largely increased use of tests on symptoms can also substantially decrease the number of deaths and daily new infections.
 The drop in daily new infections at the end of March is a direct result of the lockdown. As testing was reduced during the Easter week, we see a large drop of daily new detected infections during this time, and a short spike right after when many new infections from Easter were detected. While only the lockdown managed to reduce the estimated reproduction number to approximately 1 or below 1, increased testing lead to a lower estimated reproduction number, in particular before the lockdown. Note that the results for the reproduction number at the end of the simulation period might be inaccurate or unstable. This is due to the very low number of infected individuals that might not infect other individuals over the period of a time step.
 
In a future analysis, the cost for individual tests could be opposed to the cost of lost working days of mildly infected individuals or the additional need for hospitalization or ICU treatment to find cost-efficient mitigation strategies; see also~\cite{dorn2023common}.

\subsection{Counterfactual scenarios with increased testing instead of lockdowns}\label{AlternateReality}

\begin{figure} 
	\centering \includegraphics[width=1.0\linewidth]{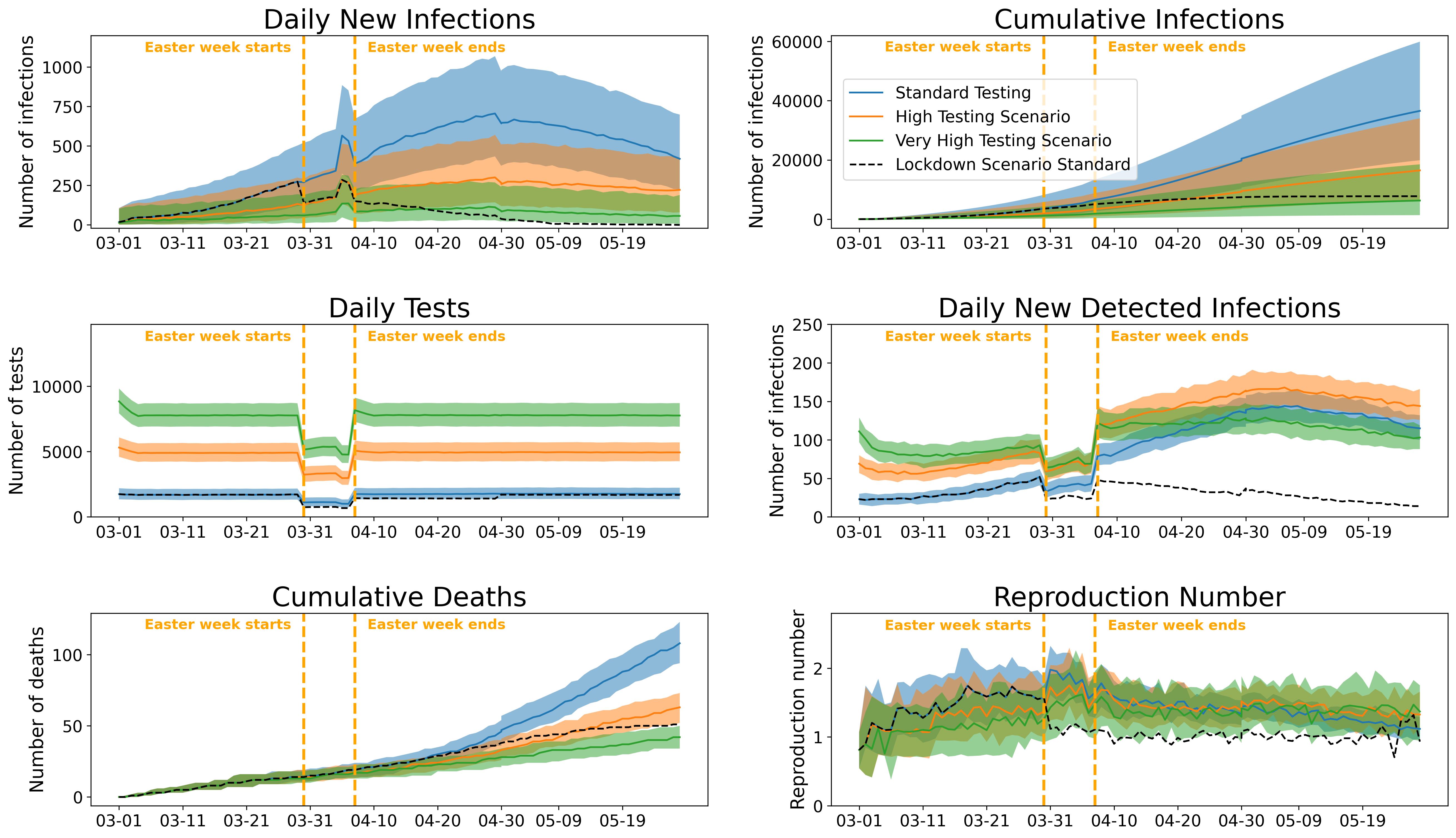} 
 	\caption{\textbf{Counterfactual scenario without lockdown but increased symptom-based testing likelihood in Brunswick between March 1st and May 30, 2021.} Standard fitted scenario without lockdown (blue) against a three- (orange) and fivefold (green) symptom-based testing rate, \SK{ where the shaded area represents the 50\% confidence interval.} During the Easter week, testing was reduced, \SK{as indicated by vertical lines. Furthermore, there are special Easter gatherings on Easter Sunday and Monday.}} 
	\label{fig:NoLockdown} 
\end{figure}

In 2020 and 2021, many countries and regions reacted to the novel Coronavirus pandemic with heavy restrictions and lockdowns to mitigate infectious disease dynamics. With the massive availability of quick and self testing capacities, the interesting research question, to which extent lockdowns could be replaced by largely increased testing, came up. In~\cite{kuhn_regional_2022}, the authors have already considered how local lockdowns and increased commuting-based testing could avoid countrywide lockdowns. In~\cref{fig:NoLockdown}, we present the results of the fitted ABM where only the lockdown has been removed from the fitting setting (blue) against a tripled (orange) and five-fold (green) testing rate to replace the lockdown. In this instance, we had no contact reduction for and after the lockdown, no reduced trips, no school closures (although staying at 50~\% reduction as before) and not a higher testing probability during the lockdown. For comparisons, we visualized the median outcome of~\cref{fig:fitting4things} with a dashed black line. Our results suggest that the lockdown has saved a lot of lives while a five-times increased testing rate, on the other hand, could have prevented even more deaths without the need for harsh restrictions.
While the lockdown manages to bring the estimated reproduction number down to approximately 1 as discussed in the previous section, a five-fold increased testing likelihood for symptomatic persons drastically reduces the estimation to almost 1, too, especially in the critical first month. This is the main reason for the low number of cumulative infections and deaths in this scenario. Note that similar findings have also been reported in the cost-benefit analysis of~\cite{eilersen_costbenefit_2020} and the testing and isolation considerations in~\cite{bandyopadhyay_testing_2022}.

\section{Discussion}\label{sec:Discussion}

ABMs allow the study of microscopic effects in infectious disease dynamics, enabling the consideration of aspects that are not possible with classical equation-based models. On the other hand, the tremendously high search space for parameter fitting and the large amount of parameter assumptions introduce uncertainties in the model outcomes. Although the model captures essential dynamics of the COVID-19 pandemic, uncertainties in key parameters, such as the viral shedding factor $s_f$ and the infection rate from viral shed $\lambda$, could influence the outcomes. Sensitivity analysis will help further quantify these parameters, which were fitted based on grid search.

In this study, we have presented a highly parametrized ABM. We have substantially reduced the parameter space by setting many parameters based on medical and biological evidence from the literature and previously validated and published models. For the remaining parameters, we were able to do an extensive grid search from which we obtained plausible values for parameters such as the initial dark figure, the symptomatic testing probability, or the ratio of symptomatic testing against untargeted testing. For the technical parameter of the infection rate from viral shed, we obtained a value of $1.596$. Unfortunately, this value cannot be compared to any data or findings from real life experiments. 
The conduction of the fitting process on multiple data sets, such as the reported case data, the ICU admissions and the reported deaths, strengthens our obtained results.

The importance of tests, as highlighted in this publication, also has to be considered in the light of parametrization. However, as we generally model rapid tests for which infections might go undetected and at the same time use higher values for mask protection than in our previous studies, we would rather expect a too pessimistic view on the effect of tests. Nevertheless, we observe for the testing strategies a large and positive impact on the mitigation of the dynamics. Note that similar findings have also been reported by, e.g.,~\cite{quilty_quarantine_2021,eilersen_costbenefit_2020,dickens_determining_2021}. Furthermore, the real-world factors like compliance rate for mask-wearing and adherence to the quarantine may vary, impacting these outcomes. Incorporating compliance data based on other statistical studies would offer a more realistic assessment of the effectiveness of these NPIs.

\MK{As a natural limitation, it is unclear to what extent the presented outcomes can be transferred to }\SKSR{regions with} \MK{settings} \SKSR{which differ in} \MK{mobility patterns, healthcare systems,} \SKSR{and} \MK{cultural practices \SKSR{from the Brunswick region}. For Germany, we, e.g., obtained similar mobility patterns and trips for the regions of Brunswick and Munich, although both are of substantially different size and located within a different geographical structure, \SKSR{compare \autoref{fig:age_side_by_side} to \autoref{fig:trip_act_side_by_side}}.} \SKSR{ For regions with substantially different contact patterns or activities, our findings might not hold true. However, our framework provides the possibility to insert generic contact and trip patterns to analyze scenarios in other regions where these patterns deviate. Enhanced testing efforts may not be as critical for controlling infection spread if the daily number of contacts is substantially smaller. A general result cannot be inferred from our findings, as these results highly depend on the contact structure (minimum, average, maximum or distribution) and on the type and intensity of contacts. These properties depend, in particular, on the cultural and socioeconomic contexts. In addition, for areas with limited healthcare infrastructure (e.g., a low number of available hospital beds), high testing rates can serve as an effective, minimally invasive intervention to curb infection spread and prevent overwhelming the healthcare system. Furthermore}, \MK{with the ABM presented here being itself a result, the openly available source code offers scientists the capability of integrating and exploring arbitrary populations and mobility patterns to infer potentially changing outcomes.}

\SK{Furthermore,} although ABMs allow a very detailed representation of the real world, simplifications and assumptions \SK{still} have to be made. While we believe to have integrated the most important features for the considered time period, additional features or extensions might be needed for the consideration of novel virus variants or long-term waning immunity. Outcomes of the presented study have been obtained in the light of a mostly immune-naive population and outcomes could be different for a highly immunized population, late-phase epidemic or endemic scenarios. \SK{Additionally}, in this study, we have considered neither the (short-term) protection against transmission, nor damped viral load courses, through vaccinations or previous infections. These effects should be integrated in an analysis for later stages of the pandemic.

Our findings suggest that quarantine length seems to play a more important role than quarantine efficiency and a minimum quarantine of 5–8 days appears to be inevitable for quarantine to have a substantial effect. While, in the literature, it is stated that even a 14-day quarantine cannot capture all uncommon or outlier infections~\cite{li_demand_2021}, several authors obtained similar findings~\cite{dickens_determining_2021,eilersen_costbenefit_2020,ashcroft_quantifying_2021,zou_trade-off_2023,wells_optimal_2021} with respect to quarantine length and differently optimized quarantining strategies. In~\cite{ashcroft_quantifying_2021}, the authors studied different test-and-release strategies. They found an diminishing relative utility of quarantine relative to a 10-day strategy and almost no additional benefit for quarantines beyond 10 days. In~\cite{eilersen_costbenefit_2020}, the authors found a five-day quarantine quite efficient, with increased effect when testing is more widespread
(compare to~\cref{fig:ParVarQuarTests}).
In~\cite{dickens_determining_2021}, the authors naturally observed increased effects for quarantines of 7, 14, and 21 days but also found that less than five percent of cases are missed with a 7-day quarantine and different test-exit strategies. The authors of~\cite{zou_epidemic_2020} found 5–10 days of quarantine most important for preventing transmissions and highlighted how test-and-release based quarantining with shorter periods can outperform 14-days quarantining with high compliance. In~\cite{wells_optimal_2021}, the authors found that testing on exit can reduce a 14-day quarantine by 50~\%.

\DK{The combination of, both, frequent tests, in particular, for symptomatic persons, and adequate quarantine shows an effective way to minimize pandemic burden in terms of case numbers. In future pandemics, where the pathogen behaves similarly to SARS-CoV-2 and no vaccine is available yet but tests are reliable and available in large amounts, we argue that this can be an effective and fast to implement way for policy-makers to keep this impact low. While these measures have been implemented to an extent in the past years in Germany, we believe that a high compliance of the population to these measures is of highest importance to successful mitigation. Thus, sensitizing the population to these measures is key. Model estimations and retrospective results such as the ones presented in this article can be used to accomplish this.

While we have not taken into account mental health and economic impact of these measures, both should be drastically reduced if a full lockdown can be avoided through these measures. Rather than having lasting mobility restrictions for the whole population over a significant time span, only a small portion of persons would have to isolate at a time. While this implication seems natural at first glance, a more detailed behavioral model that includes compliance and mental health would need to be integrated into the whole ABM to analyze NPIs in this regard. The authors of~\cite{zozmann_autonomous_2024} discuss autonomous and policy-induced behavior change during the COVID-19 pandemic, and explain the importance of these aspects in epidemiological models. This field is not explored nearly enough, and will be part of future work.
We stress that a reevaluation of NPIs is necessary for a new pathogen, in particular, when its properties vastly differ from the ones considered here, or for different environments, where social infrastructure or cultural habits differ from the setting investigated here. Also, with vaccines being available in a large amount, the importance of testing and isolation can potentially decrease, depending on the political goal. This has not been studied in this article, however.
}

\section{Conclusion}

In this study, we presented an efficiently implemented and parallelized ABM that can be executed with hundreds of thousands to millions of agents on a consumer laptop. Furthermore, it allows the use of a hybrid, shared and distributed memory parallelism to compute large numbers of ensemble results on a supercomputing facility. This capability is essential for handling the computational complexity of ABMs, especially when exploring \SK{a} wide parameter space or performing sensitivity analysis.

On the application side, we fitted our ABM for the period of March to May 2021 in the Brunswick region in Germany. We can well reproduce the cumulative deaths, the detected number of infections and, except for two smaller drops in ICU admissions, also ICU admissions. We also correctly reproduce the drop in detected new infections over the Easter period in 2021.

Furthermore, to show the effects of different testing and isolation strategies and the sensitivity of the model with respect to (combined) parameter variations, we provided different heat maps for the aggregated outcomes or endpoints. We consider cumulative infections, deaths, maximum hospitalized patients and maximum daily infections over the three-month simulation interval when varying one parameter against another.

With the limitations as given in the discussion section, we demonstrated how a minimal-invasive but significantly increased symptom-based testing strategy could have replaced stricter interventions of a lockdown. However, the success of such intervention is highly dependent on population compliance. Ensuring public cooperation and availability of testing resources is crucial for maximizing its impact. We therefore consider it of utmost importance to sensitize the population to regular testing and to make sufficient tests available at an affordable price. 

We furthermore studied the combined effects of symptom-based and untargeted, symptom-independent testing strategies\SK{,} as well as quarantine length and efficiency. We observed that with increased testing of symptomatic persons, the testing of nonsymptomatic persons plays a higher role, leading to a larger reduction in all considered endpoints. On the other hand, with weak symptomatic control, i.e., with a low likelihood to test symptomatic persons, untargeted testing was observed to be mostly inefficient. Furthermore, quarantine length seems to play a more important role than quarantine efficiency and a minimum quarantine of 5–8 days appears to be inevitable for quarantine to have a substantial effect. This, of course, is in line with high viral loads that have been observed at the beginning of the infection. Similar findings have been discussed in the previous section. Finally, we also observed how high symptomatic testing rates leverage quarantining effects\SK{,} such that even short quarantining can have a large effect on mitigation. 

The flexibility of this ABM also allows for adaptation to study other infectious diseases or apply it to different geographic regions with varying epidemiological factors. Future studies could leverage this model to explore the impacts of emerging variants or long-term immunity, offering insights into public health strategies beyond COVID-19.

\section*{Acknowledgments}
The authors gratefully acknowledge computing time on the supercomputer JURECA~\cite{julich_supercomputing_centre_jureca_2021} at Forschungszentrum Jülich under grant HPC4EPI. DK, MMH, and MJK have received funding from the Initiative and Networking Fund of the Helmholtz Association (grant agreement number KA1-Co-08, Project LOKI-Pandemics).
The authors DA, CG, AS, and MJK have received funding by the German Federal Ministry for Digital and Transport under grant agreement FKZ19F2211A (Project PANDEMOS), KN \DKSR{and MMH have} received funding by the German Federal Ministry for Digital and Transport under grant agreement FKZ19F2211B (Project PANDEMOS). SK has performed his work as part of the Helmholtz School for Data Science in Life, Earth and Energy (HDS-LEE) and received funding from the Helmholtz Association of German Research Centres. The funders had no role in study design, data collection and analysis, decision to publish, or preparation of the manuscript.

\section*{Competing interests}
The authors declare to not have any competing interests.

\section*{Data availability}

The code is publicly available on GitHub under \url{https://github.com/SciCompMod/memilio}, the actual simulation code is to be found on branch \textit{abm\_paper\_test\_bs}. Mobility data provided and used by this study is available on~\cite{schengen_2024_13318436}, contact data is included in the repository itself. Given this data, reproduction of our results is possible. 

\section*{Author Contributions}

    \noindent\textbf{Conceptualization:} Daniel Abele, Martin Kühn, Michael Meyer-Hermann\\
    \noindent\textbf{Data Curation:} David Kerkmann, Sascha Korf, Alain Schengen, Carlotta Gerstein\\
    \noindent\textbf{Formal Analysis:} David Kerkmann, Sascha Korf, Khoa Nguyen, Daniel Abele, Martin Kühn\\
    \noindent\textbf{Funding Acquisition:} Martin Kühn, Michael Meyer-Hermann, Achim Basermann\\
    \noindent\textbf{Investigation:} David Kerkmann, Sascha Korf, Khoa Nguyen, Martin Kühn, Michael Meyer-Hermann\\
    \noindent\textbf{Methodology:} David Kerkmann, Sascha Korf, Khoa Nguyen, Daniel Abele, Alain Schengen, Carlotta Gerstein, Jens Henrik Göbbert, Martin Kühn, Michael Meyer-Hermann\\
    \noindent\textbf{Project Administration:} Achim Basermann, Martin Kühn, Michael Meyer-Hermann\\
    \noindent\textbf{Resources:} Achim Basermann, Martin Kühn, Michael Meyer-Hermann\\
    \noindent\textbf{Software:} David Kerkmann, Sascha Korf, Khoa Nguyen, Daniel Abele, Carlotta Gerstein, Martin Kühn\\
    \noindent\textbf{Supervision:} Martin Kühn, Michael Meyer-Hermann \\
    \noindent\textbf{Validation:} All authors \\
    \noindent\textbf{Visualization:} David Kerkmann, Sascha Korf, Khoa Nguyen, Carlotta Gerstein\\
    \noindent\textbf{Writing – Original Draft:} David Kerkmann, Sascha Korf, Khoa Nguyen, Martin Kühn\\
    \noindent\textbf{Writing – Review \& Editing:} All authors

\bibliographystyle{elsarticle-num}
\bibliography{refs}

\appendix

\section{Comparison of the Munich and Brunswick region}

\SKSR{Since data for the urban city of Munich in Germany was readily available, we compared certain trip patterns between Munich and Brunswick to assess whether the findings could be generalized to other urban environments within Germany.
Here, we again use the base population of people who live in the respective city plus the incommuters and their households.
While the results show a notable similarity between the two cities this comparison is not exhaustive. 
It provides only an initial insight, as Munich and Brunswick share several key characteristics. 
However, these findings may not fully apply to smaller cities or non-German urban environments with distinct characteristics, which could exhibit different trip behaviors due to varying socio-economic and cultural factors.}

\begin{figure}[h!]
    \centering
    \includegraphics[width=0.95\linewidth]{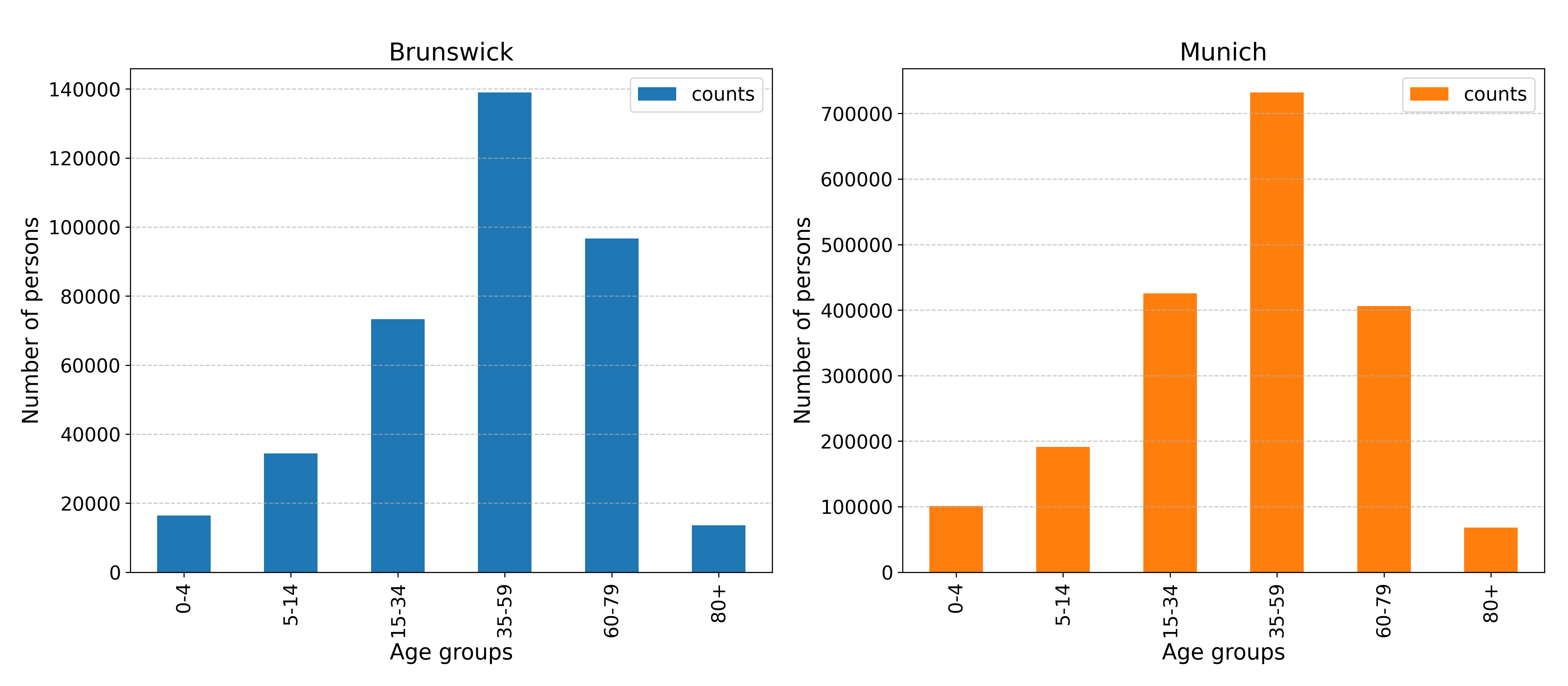}
    \caption{\textbf{Age distribution in the synthetic population of the regions of Brunswick and Munich.}}
    \label{fig:age_side_by_side}
\end{figure}

\begin{figure}[h!]
    \centering
    \includegraphics[width=0.95\linewidth]{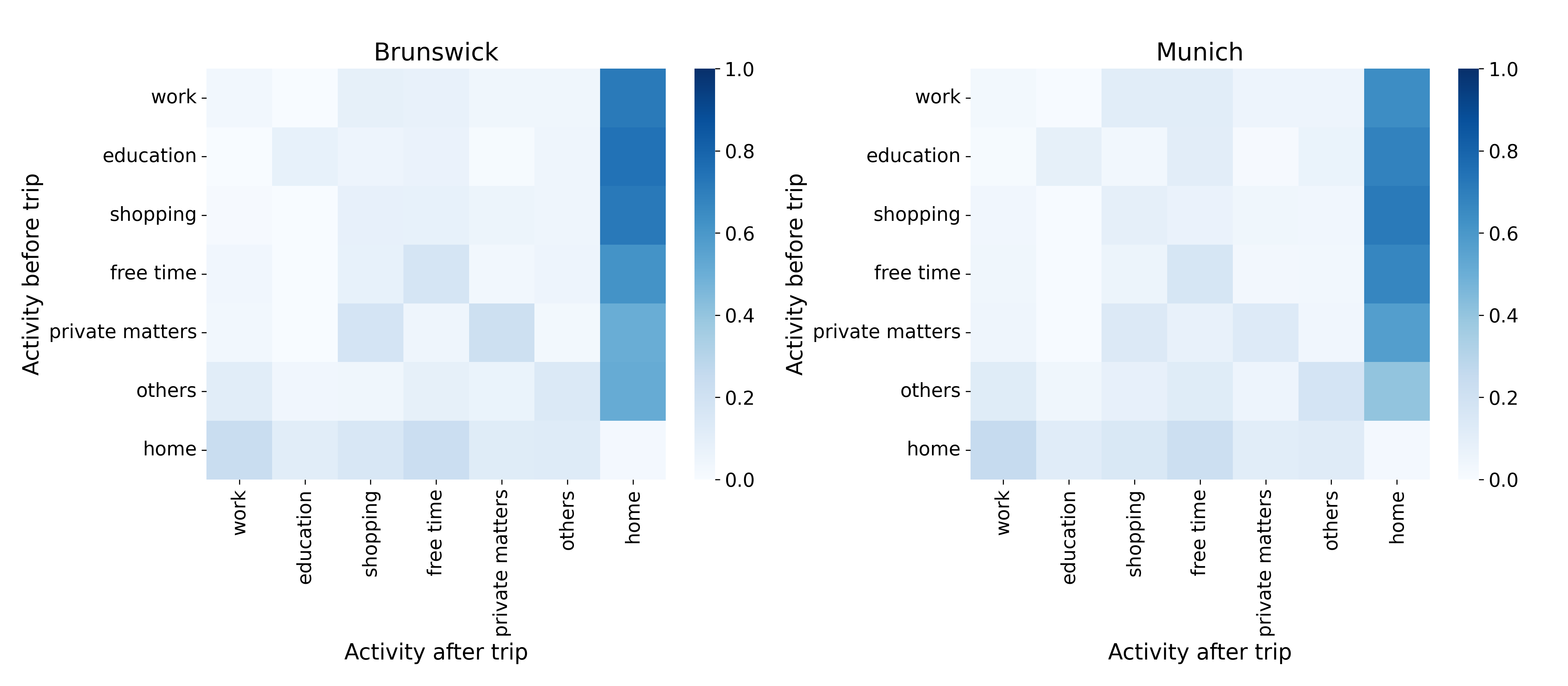}
    \caption{\textbf{Heatmap of trip frequency of origin-destination type for the locations in the synthetic population of the regions of Brunswick and Munich.}}
    \label{fig:od_side_by_side}
\end{figure}

\begin{figure}[h!]
    \centering
    \includegraphics[width=1.0\linewidth]{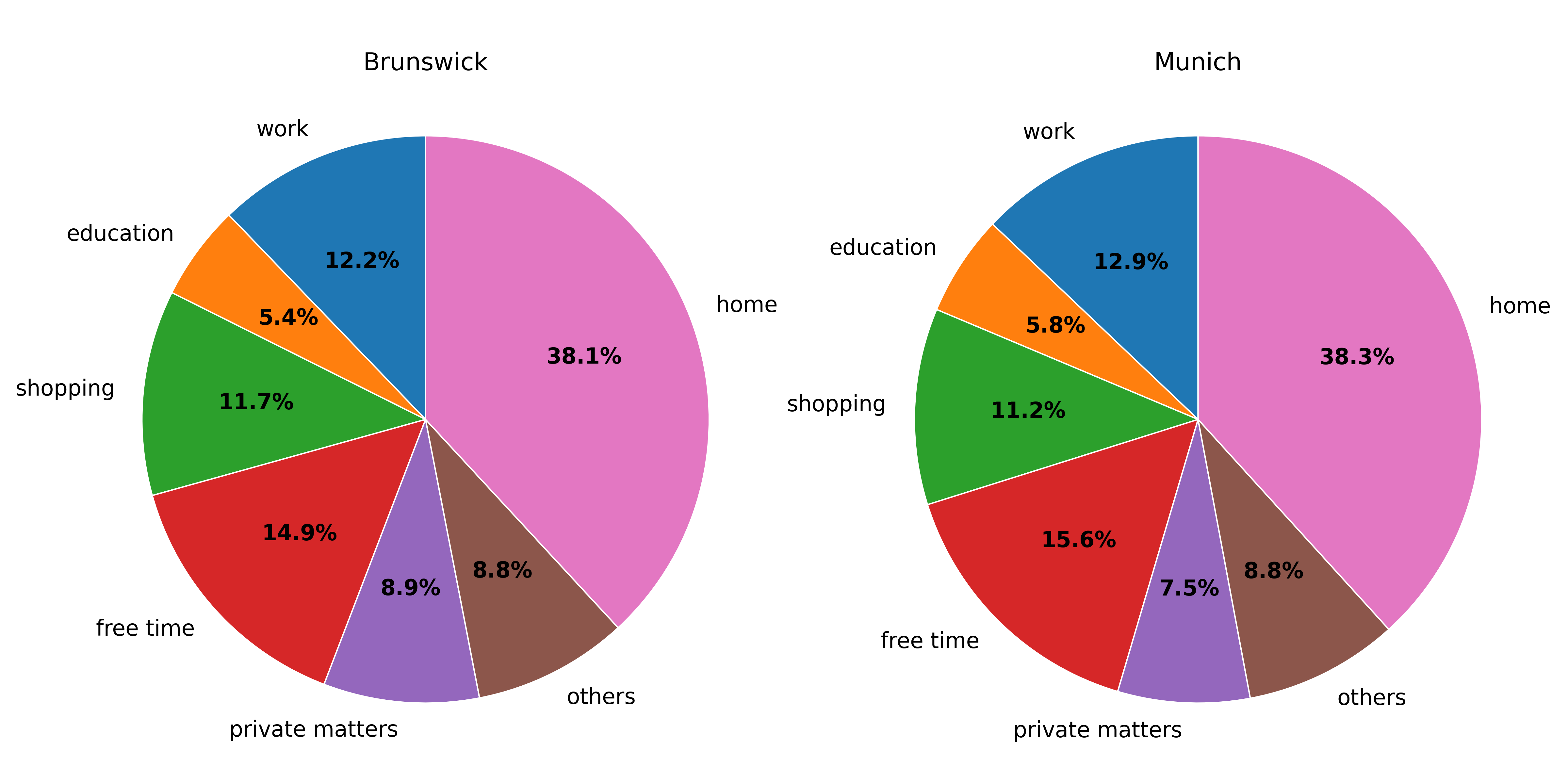}
    \caption{\textbf{Trip activity distribution for the overall population of in the synthetic population of the regions of Brunswick and Munich.}}
    \label{fig:trips_rel_side_by_side}
\end{figure}

\begin{figure}[h!]
    \centering
    \includegraphics[width=0.95\linewidth]{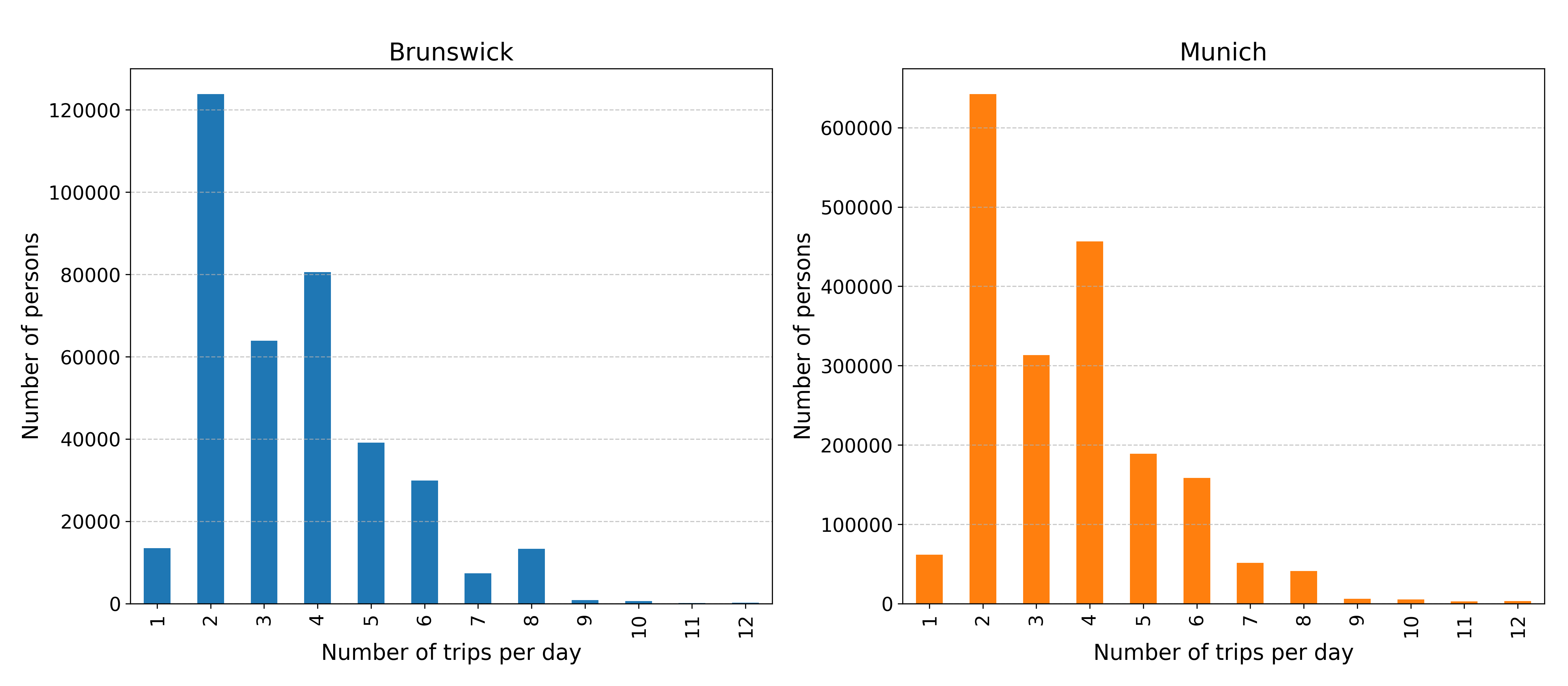}
    \caption{\textbf{Number of trips per person in the synthetic population of the regions of Brunswick and Munich.}}
    \label{fig:trips_per_person_side_by_side}
\end{figure}

\begin{figure}[h!]
    \centering
    \includegraphics[width=1.0\linewidth]{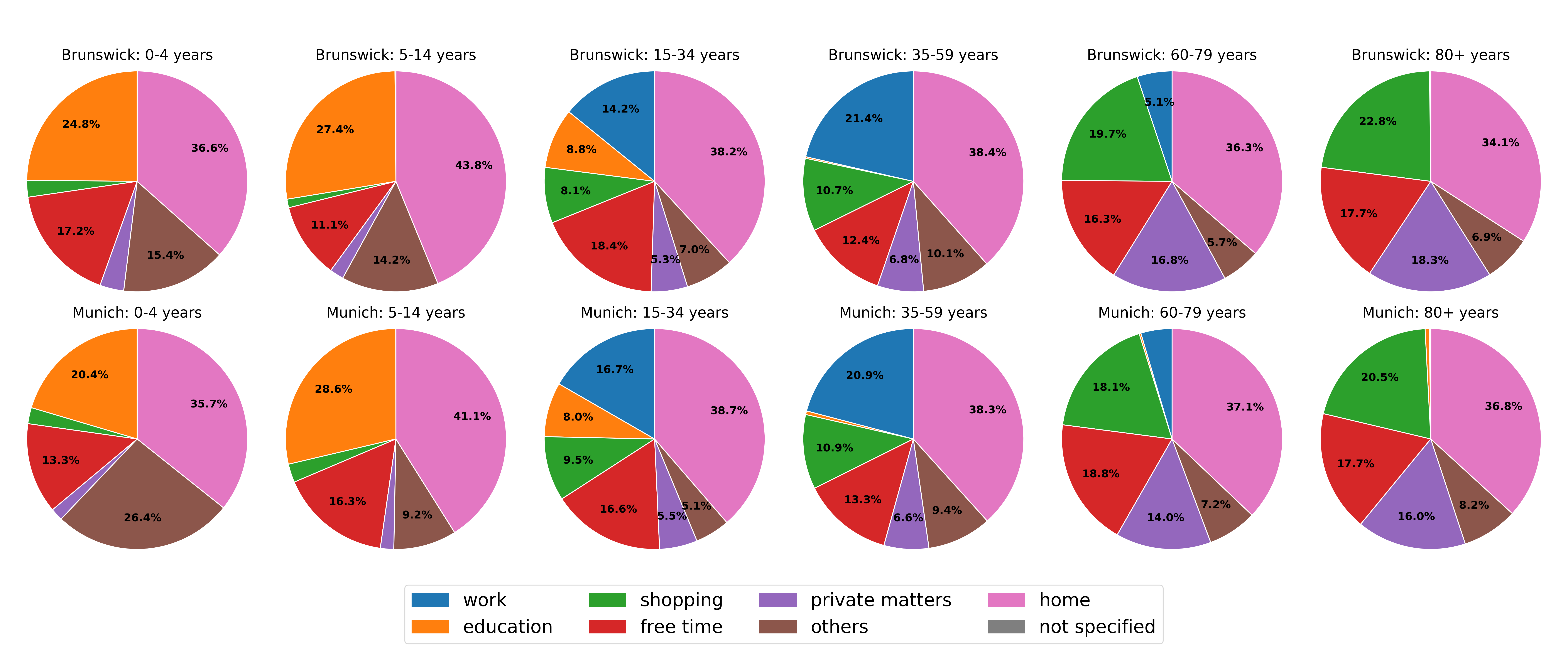}
    \caption{\textbf{Trip activity distribution resolved by age groups in the synthetic population of the regions of Brunswick and Munich.}}
    \label{fig:trip_act_side_by_side}
\end{figure}

\MKSR{As~\cref{fig:age_side_by_side} illustrates, we have similar age distributions in the synthetic populations of the regions of Brunswick and Munich. In~\cref{fig:od_side_by_side}, the heatmap shows the trip frequency between the different location types. We see that both regions share similar trip frequencies between the different location types. Especially, both regions share that most trips go to home location. A similar insight is obtained with~\cref{fig:trips_rel_side_by_side}. The comparison shows that, relatively, both regions show a very similar distribution of the trip purpose for the overall population.

In~\cref{fig:trips_per_person_side_by_side}, the comparison shows that, relatively, the number of persons for each number of trips is very similar for both regions. Eventually,~\cref{fig:trip_act_side_by_side} also shows that the age-resolved trip data is similar between the two different regions.}

\end{document}